\documentclass[12pt,a4paper,final]{iopart}

\usepackage{iopams}
\usepackage{graphicx}
\usepackage{hyperref}

\begin{document}

\title[]{Long-Term Postmerger Simulations of Binary Neutron Star Coalescence: Formation of Toroidal Remnants and Gravitational Wave Afterglow}




\author[cor1]{Xuefeng Zhang}
\address{TianQin Reseach Center for Gravitational Physics, \\ School of Physics and Astronomy, Sun Yat-sen University, Zhuhai, China}
\ead{zhangxf38@sysu.edu.cn}

\author[]{Zhoujian Cao}
\address{Department of Astronomy, Beijing Normal University, Beijing, China}
\eads{\mailto{zjcao@amt.ac.cn}}


\begin{abstract}
It has been estimated that a significant proportion of binary neutron star merger events produce long-lived massive remnants supported by differential rotation and subject to rotational instabilities. To examine formation and oscillation of rapidly rotating neutron stars (NS) after merger, we present an exploratory study of fully general-relativistic hydrodynamic simulations using the public code {\tt Einstein Toolkit}. The attention is focused on qualitative aspects of long-term postmerger evolution. As simplified test models, we use a moderately stiff $\Gamma=2$ ideal-fluid equation of state and unmagnetized irrotational equal-mass binaries with three masses well below the threshold for prompt collapse. Our high resolution simulations generate postmerger ``ringdown'' gravitational wave (GW) signals of 170 ms, sustained by rotating massive NS remnants without collapsing to black holes. We observe that the high-density double-core structure inside the remnants gradually turns into a quasi-axisymmetric toroidal shape. It oscillates in a quasi-periodic manner and shrinks in size due to gravitational radiation. In the GW spectrograms, dominant double peaks persist throughout the postmerger simulations and slowly drift to higher frequencies. A new low-frequency peak emerges at about 100 ms after merger, owing to the growth of GW-driven unstable oscillation modes. The long-term effect of grid resolution is also investigated using the same initial model. Moreover, we comment on physical conditions that are favorable for the transient toroidal configuration to form, and discuss implication of our findings on future GW observation targeting rapidly rotating NSs.
\end{abstract}


\submitto{} 

\section{Introduction} \label{sec:intro}
Neutron stars (NS) can generate gravitational waves (GW) in a variety of ways \cite{Andersson11}.  Particularly, binary neutron star (BNS) mergers are among the most promising GW sources, likely to be detected in the near future and possibly accompanied by electromagnetic counterparts. A major distinction of BNS mergers from binary black hole (BBH) mergers lies in the former's much more complicated postmerger processes. Following merger, there can be roughly three outcomes \cite{Faber12,Baiotti17}: for high-mass binaries, (1) a BH from prompt gravitational collapse; for low and medium-mass binaries, (2) a stable NS, or (3) a massive NS that is supported by fast rotation and eventually collapses to a BH due to angular momentum dissipation. Based on current estimation, NS binaries with total masses $M\leq M_{thr}$ ($\lesssim 2.9 M_\odot$, depending on EOSs) are expected to produce hot massive remnants that may survive from a few tens of milliseconds up to hours \cite{Ravi14}. Given that most observed binary systems have NS masses within a very narrow range of $1.33\pm 0.09 M_\odot$ \cite{Ozel16}, a significant fraction of merger events are expected to form massive NS remnants (e.g., \cite{Gao16,Piro17}). The remnant lifetime, i.e., the time before collapse, depends nonlinearly on a number of factors, such as the equation of state (EOS), mass ratios, magnetic fields, viscosity, etc. Hence it is difficult to estimate accurately and poses challenges on theoretical models and numerical simulations.

So far numerical relativity (NR) is the only option to treat BNS mergers with precision (see recent reviews \cite{Duez10,Faber12,Baiotti17,Paschalidis17}). Most BNS simulations up to date have focused on relatively short periods after merger and comparing merger dynamics among several EOS candidates. Current published postmerger simulations mostly last $\sim 20$ ms and not longer than $\sim 50$ ms, barring one rare exception of 130 ms \cite{Rezzolla10} to our best knowledge. The short duration is partly due to high cost of NR simulations. Also in the case of prompt collapse into BHs or rapid decay of postmerger GW signals, running long simulations is simply not motivated, at least for the purpose of GW detection. However, with certain EOSs (e.g., $\Gamma=2$ ideal fluid, GNH3, H4), binary mergers can indeed lead to systematically stronger postmerger GW emission than other EOSs \cite{Takami15,Rezzolla16}. These high-frequency (1-4 kHz) GW radiations, produced by non-axisymmetric deformation of rapidly spinning self-gravitating NSs, may potentially become detectable by future ground-based detectors, such as the Einstein Telescope \cite{Punturo10a,Punturo10b}. The signals rely heavily on NS physics and will be rich in information to constrain the nuclear EOS \cite{Shibata05c}. Hence it would be worthwhile to look into postmerger evolution of extended lengths and examine associated GW emission.

Rapidly differentially rotating NSs naturally arise in BNS mergers, which may trigger the onset of rotational instabilities. From both theoretical and observational viewpoints, there has been long-standing interest in rotating relativistic stars subject to various instabilities \cite{Stergioulas03,Paschalidis16}. By the prevalent method, to build a differentially rotating NS in equilibrium, one needs to put in man-made rotation profiles that may or may not be realistic. For instance, the so-called $j$-constant rotation law has been commonly used, nevertheless its realism being questioned \cite{Hanauske16}. More physically relevant rotation law should be obtained through numerical simulations of the formation of rotating NSs from stellar collapse or BNS merger.

From a rotating NS model, to determine the dynamical instability and subsequent evolution requires NR simulations. So far, it has been shown by example that differentially rotating hypermassive NS models can be dynamically stable \cite{Baumgarte00}. Particularly if the rotation is strong enough, the central rest-mass density of NSs does not need to coincide with the maximum density \cite{Baumgarte00}. Thereby one can construct NS models of a torus-like shape (high density region forming a torus) that are stable on dynamical timescales \cite{Duez04,Duez06,Baiotti07,DePietri14,Loffler15,Gondek17}. Furthermore, the growth and saturation of the bar-mode ($m=2$) instability have also been demonstrated \cite{Shibata00b}. They allow NSs to sustain bar-shaped deformation for many accompanying quasi-periodic GW cycles \cite{Lai95}. Similar mechanisms are expected to perpetuate GW emission in massive NSs from BNS mergers. In addition, secularly unstable oscillation modes of isolated NSs have also been considered as potential astrophysical sources for GW search (see, e.g., \cite{Andersson98,Owen02,Andersson03,Bondarescu07,Bondarescu09,Owen09,Kruger10,Kastaun10}), and their detection can provide extra constraints to the NS microphysics. But it is important to first clarify under what circumstances these modes can be excited. By running long-term postmerger simulations, we are trying to build a link between possible secular dynamics of single NSs and the physical scenario of BNS coalescence.

In this paper, we consider idealized, and somewhat extreme, situations where one has persistent postmerger GW emission (driven by double-core structure), and investigate long-term behavior of the merger remnants on a 170 ms timescale. This time frame is when the remnants have reached hydrodynamic quasi-equilibrium for long enough so that one may collect and extrapolate information of secular evolution on longer timescales. Unfortunately, due to poorly known NS physics and high cost of long-term simulations using high resolution, we have limited ourselves to simplified binary configurations for a proof-of-concept case study. First, we have picked $\Gamma=2$ ideal fluid ($\Gamma$-law) EOS as a crude representative of stiff nuclear EOSs. Historically, this simple EOS has been widely used in numerous published works involving NSs. Second, as a first-step approximation to more realistic scenarios, we assume unmagnetized equal-mass binaries with no individual spins, no physical viscosity, no radiative transport, and no crust in our models. This basic setup is fairly common in the literature, since the effects of physical viscosity, magnetic fields, and radiative transport are expected to act on longer timescales \cite{Paschalidis17} and not to cause significant changes in GW signals (see, e.g., \cite{Shibata05a,Giacomazzo11,Palenzuela15,Ciolfi17}). Despite the model being quite idealized, we consider it sufficient for qualitative exploration and plausible on the moderately long timescale of 170 ms. The ``clean'' setup also helps us to focus on understanding the underlying physics and basic principles, as well as setting up a benchmark baseline, before delving into more realistic and complex situations.

Long-term BNS postmerger simulation has been attempted before, notably in \cite{Rezzolla10} (see the Appendix therein and also discussions in \cite{Baiotti08}) under the same physical setup described above. The simulation keeps track of a hypermassive NS for 120 ms before it collapses to a BH. The result was presented partly to showcase the accuracy and stability of the numerical code ({\tt Whisky}) over a long period. However, not much was discussed on long-term dynamics. Here we will instead focus on these aspects and push the length of simulations even further. Other examples featuring simulating long-lived NSs up to $\sim 50$ ms after merger can be found in, e.g., \cite{Giacomazzo13,Bernuzzi16,Radice16,Ciolfi17}.

We take running long-term simulations as a stringent test of the public-available {\tt Einstein Toolkit} \cite{ETK}, which allows for independent confirmation and repetition of the results. We hope to demonstrate the degree of feasibility to perform long-term stable simulations with the {\tt Einstein Toolkit} on a qualitatively accurate level. Nevertheless, it must be emphasized that the purpose of this study is not to create long waveforms with reliable phase, which would be impractical with current numerical schemes, but to identify robust qualitative phenomena and to serve as pointer for future simulations. This point of view also applies to our assessment of the influence of grid resolution.

The paper is organized as follows. In Sec. \ref{sec:setup}, the setup for numerical simulations and the description of physical models are presented. The main body of the paper is devoted to two main aspects of the BNS time evolution. First in Sec. \ref{sec:masses}, we will carry out three runs that only differ in NS baryon masses. We will look into oscillatory motion of long-lived rotating massive NS remnants, and explore GW spectral properties in the quasi-stationary phase that last from $\sim$20 to 170 ms after merger. Then in Sec. \ref{sec:resltn}, we pin down one initial data model ($M_b=1.445 M_\odot$) and examine the qualitative effect of spatial resolution on the simulations using five finest grid sizes ranging from $\rmd x=9/64$ to $18/64$. In these two sections, we also discuss and identify challenges and complications of performing long-term simulations. The paper concludes in Sec. \ref{sec:conclu}. Throughout the paper, we employ geometrized units $G=c=M_\odot=1$, and $t$ denotes the coordinate time.


\section{Numerical Setup and Models} \label{sec:setup}

To capture the highly relativistic dynamics of BNS mergers starting from late inspiral phases, one needs to solve the Einstein's equations for dynamical spacetime, coupled with the fluid equations describing NS matter. Our simulations are implemented using the {\tt Einstein Toolkit} \cite{ETK} (the 11th release {\tt ET\_2015\_05} ``Hilbert'' ), which is a collection of public available, open-source, and community-driven NR codes first released in 2010. The program is built on the {\tt Cactus} infrastructure \cite{Cactus} for high-performance computing in 3 dimensions. The development of the {\tt Einstein Toolkit} has been well documented, and more details can be found in, e.g., \cite{Loffler12,Mosta14}.

Using the {\tt Einstein Toolkit} in BNS studies has been demonstrated by the Parma group \cite{DePietri16,Maione16,Feo16}. An exemplary set of simulation files is also available on the {\tt Einstein Toolkit} website \cite{ETK}. Our implementation is built on these examples with necessary modifications to the parameter files. Here we summarize the main aspects.

(1) The outer boundary of the Cartesian grid is set at $d=540\simeq 798$ km from the origin. We use 7 levels of adaptive mesh refinement (AMR) with nested ``moving boxes'' encasing each NS. The grid hierarchy is implemented by the {\tt Carpet} code \cite{Carpet}. Our high resolution runs have the grid spacing $\rmd x=9/2^6=0.140625\simeq208\ \mathrm{m}$ on the finest level, which is slightly smaller than $\rmd x=0.15$ that has been commonly used for accurate evolution. We adopt this higher resolution as an attempt to reduce computational error accumulated over extended simulation time. However, doing so makes the computational cost grow drastically both in memory usage and computation time. The chosen resolution appears to be a balanced compromise.

(2) To reduce computational cost, we impose $\pi$-rotational symmetry around the $z$-axis, hence ignoring one-armed spiral ($m=1$) modes which undercut persistence of the bar-mode ($m=2$) deformation \cite{Baiotti07,Manca07,Radice16,Lehner16}. A reflection symmetry across the orbital plane $z=0$ is also applied to chop off another half of the computational domain.

(3) The spacetime metric is evolved by the {\tt McLachlan} code \cite{McLachlan}. It implements the 3+1 dimensional BSSN formalism \cite{Shibata95,Baumgarte99} and high-order finite-difference techniques, as well as Kreiss-Oliger dissipation for suppressing high-frequency numerical noises. For time integration, we use the fourth-order Runge-Kutta method and set the Courant factor to 0.4, a slightly large value to further trim down computational cost.

(4) The hydrodynamic evolution is handled by GRHydro \cite{Loffler12,Mosta14}, a module derived from the public {\tt Whisky} code \cite{Whisky}. It employs a high resolution shock capturing (HRSC) scheme. Our simulations utilize a combination of the piecewise parabolic reconstruction method (PPM) and the Harten-Lax-van Leer-Einfeldt (HLLE) Riemann solver.

NR simulations over one hundred milliseconds are computationally expensive and time-consuming. One high resolution simulation with $\rmd x= 0.140625$ occupied 144 CPU cores for about 25 days, and generated time evolution and waveform of 180 ms. The duration alone puts the {\tt Einstein Toolkit} on a demanding test.

\subsection{Equation of state}

A biggest uncertainty in BNS simulations is the EOS because of poorly known microphysics under extreme conditions. For time evolution, we have used an ideal fluid ($\Gamma$-law) EOS:
\begin{equation*}
    P = (\Gamma - 1) \rho \epsilon, \qquad \Gamma=2.
\end{equation*}
The equation relates the pressure $P$ to the rest-mass density $\rho$ and the specific internal energy density $\epsilon$. It allows for increase of internal energy and entropy via shock heating. Also we adopt the adiabatic exponent $\Gamma=2$ as a simple qualitative approximation to moderately stiff nuclear EOSs. For the initial data in cold hydrodynamic equilibrium, we assume a simple polytropic EOS:
\begin{equation*}
    P = K \rho^\Gamma,
\end{equation*}
with the polytropic coefficient $K=123.6$. The corresponding maximum rest mass of non-rotating NSs can reach $2M_\odot$ \cite{Cook94}. The $\Gamma=2$ EOS has been widely used as a test model in various studies concerning NSs (e.g., \cite{Shibata00a,Baiotti08} on BNS mergers). Its simplicity helps to isolate effects not associated with microphysics. For a comparison with other nuclear EOSs, see, e.g., FIG.1 in \cite{Takami15}.

\subsection{Initial data}

\begin{table}
 \caption{\label{tab_model} Initial data models and their parameters. The list includes the baryon mass $M_b$ of each star, the total gravitational mass $M_\mathrm{ADM}$, the angular momentum $J$, the orbital angular velocity $\Omega_0$, the NS radius $R_y$ along the initial orbital velocity ($y$-direction), and the maximum rest-mass density $\rho_{max}$. All the models have an initial coordinate separation of 45 km between stellar centers. }
 \begin{indented}
 \lineup
 \item[]\begin{tabular}{cccccc}
 \br
   $M_b$($M_\odot$) & $M_\mathrm{ADM}$($M_\odot$) & $J$($GM_\odot^2/c$) & $\Omega_0$(rad/ms) & $R_y$(km) & $\rho_{max}$($c^6/G^3M_\odot^2$) \cr
 \mr
   1.355 & 2.5393 & 6.705 & 1.741 & 13.95 & $6.568\times 10^{-4}$  \cr
   1.399 & 2.6150 & 7.037 & 1.760 & 13.75 & $6.969\times 10^{-4}$  \cr
   1.445 & 2.6938 & 7.391 & 1.779 & 13.54 & $7.415\times 10^{-4}$  \cr
  \br
 \end{tabular}
 \end{indented}
\end{table}

To generate initial data for irrotational equal-mass binaries in quasi-circular orbits, we use the publicly available multi-domain spectral code {\tt LORENE} \cite{Lorene,Gourgoulhon01,Taniguchi02}. Three initial configurations are considered where an individual star carries baryon mass $M_b=1.355$, $1.399$, $1.445 M_\odot$. All the binaries evolve from an initial coordinate separation of 45 km between stellar centers. Their properties are summarized in Table \ref{tab_model}. Given the high resolution $\rmd x=0.140625\simeq 208$ m, the initial neutron star is resolved by $\sim 130$ grid points along the diameter.

Since we are interested in long-lived massive NS remnants, the initial binary masses have been hand-picked for relatively low values to avoid prompt collapse to BHs after mergers. The threshold for prompt collapse is $2M_b\sim 3.4 M_\odot$ \cite{Shibata03,Baiotti08}. Also we recall that in \cite{Rezzolla10}, a $\Gamma=2$ equal-mass binary model with $M_b=1.456 M_\odot$ results in a very much delayed collapse in 120 ms after merger. Hence the remnants from our chosen low-mass binaries should have lifetimes over 100 ms.

\subsection{GW signals}

The GW signals are extracted at various fixed coordinate radii using the Newman-Penrose formalism, which evaluates the Weyl scalar $\psi_4$ (via the module {\tt WeylScal4}). In the radiation zone, the GW strain is related to the scalar $\psi_4$ by (see, e.g., \cite{Ruiz08})
\begin{equation}
    \psi_4 = \ddot{h}_+ - \rmi \ddot{h}_\times.
\end{equation}
In this work, we only consider the most dominant $l=2=m$ mode by decomposition in terms of $s=-2$ spin-weighted spherical harmonics. The power spectral density (PSD) of the signals is defined by
\begin{equation}
    \tilde{h}(f) = \sqrt{\frac{|\tilde{h}_+(f)|^2+|\tilde{h}_\times(f)|^2}{2}},
\end{equation}
where $\tilde{h}_{+,\times}(f)$ are the Fourier transformation of $h_{+,\times}(t)$. Without including physical viscosity, magnetic fields, and thermal radiation, we assume GW emission as the dominant dissipation over the timescale of our interest.


\section{Evolution with Different Masses} \label{sec:masses}

We have computed the time evolution from the three equal-mass binary initial models presented in the previous section. They are evolved using the high resolution $\rmd x=9/64=0.140625$. The discussion will focus on qualitative behavior of the long-term postmerger evolution and highlight most salient properties.

\subsection{Remnant dynamics and morphology}

\begin{figure}[htbp!]
 \centering
 \mbox{
 \includegraphics[width=0.45\textwidth]{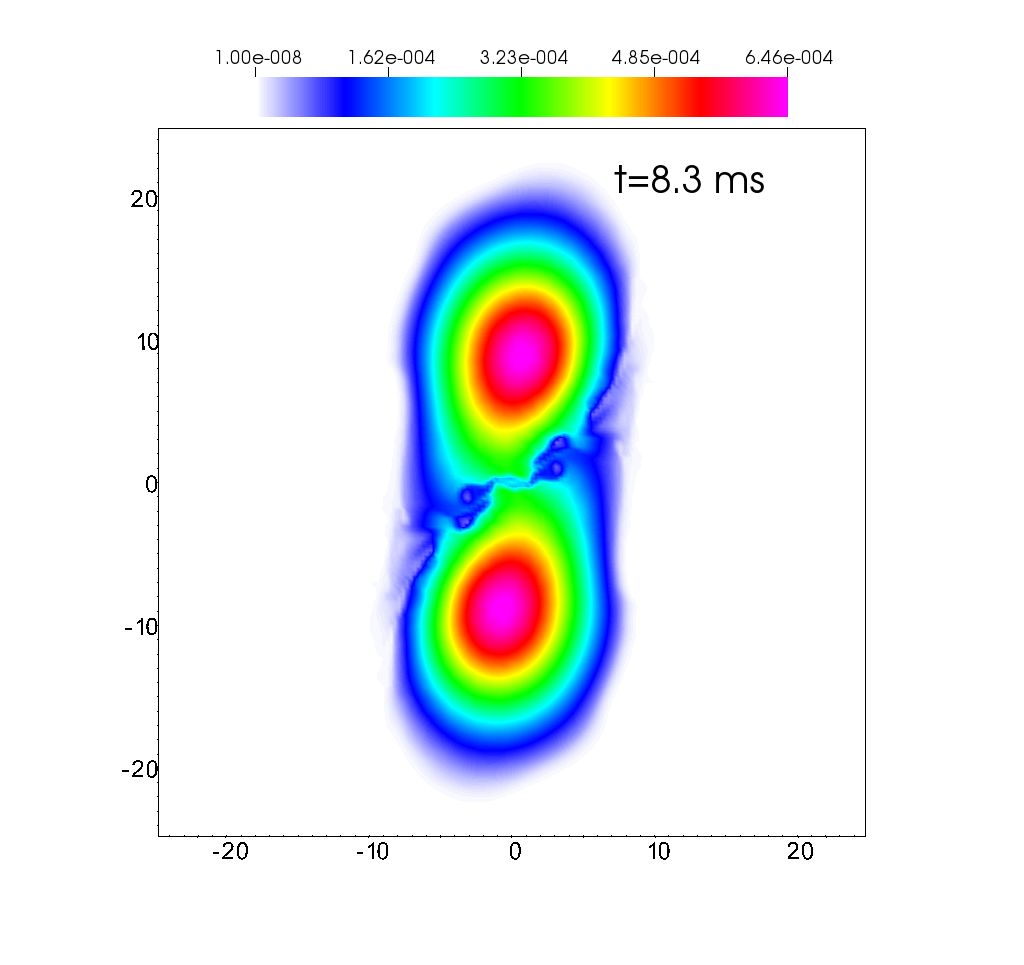}
 \includegraphics[width=0.45\textwidth]{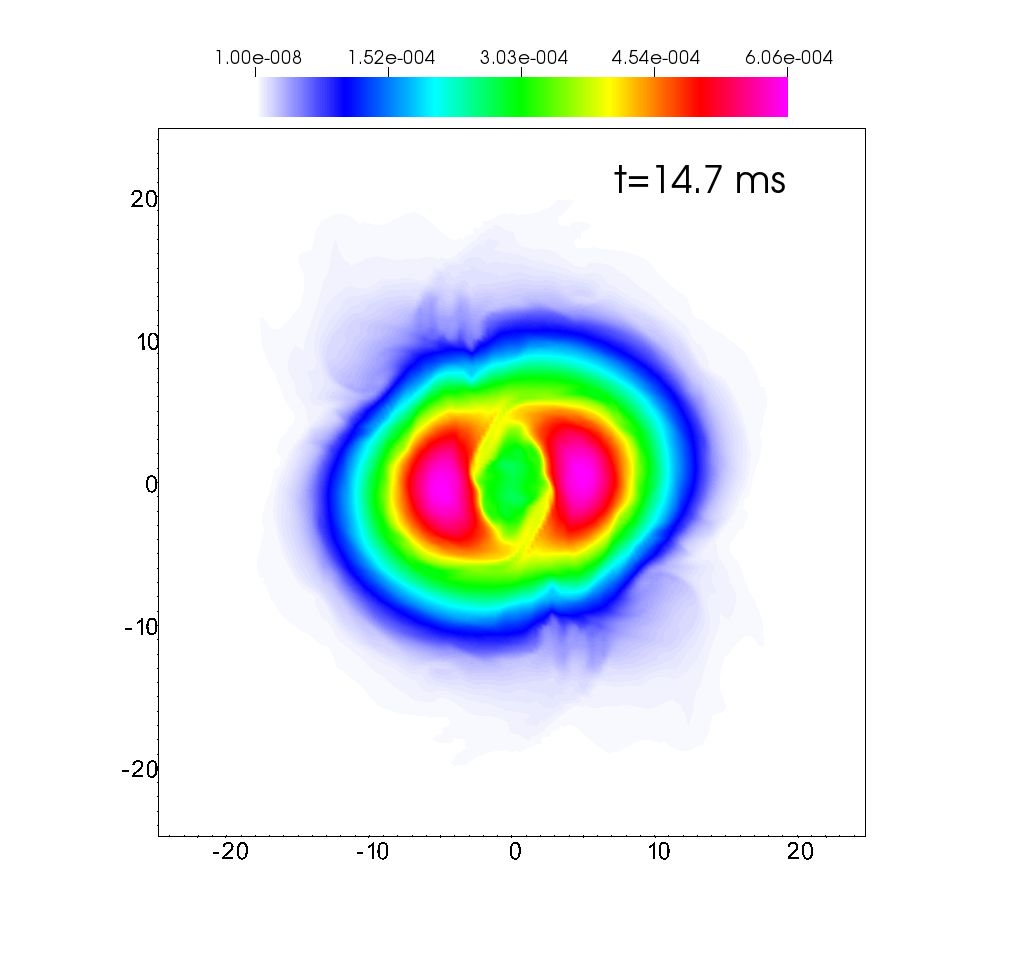}
 }
 \mbox{
 \includegraphics[width=0.45\textwidth]{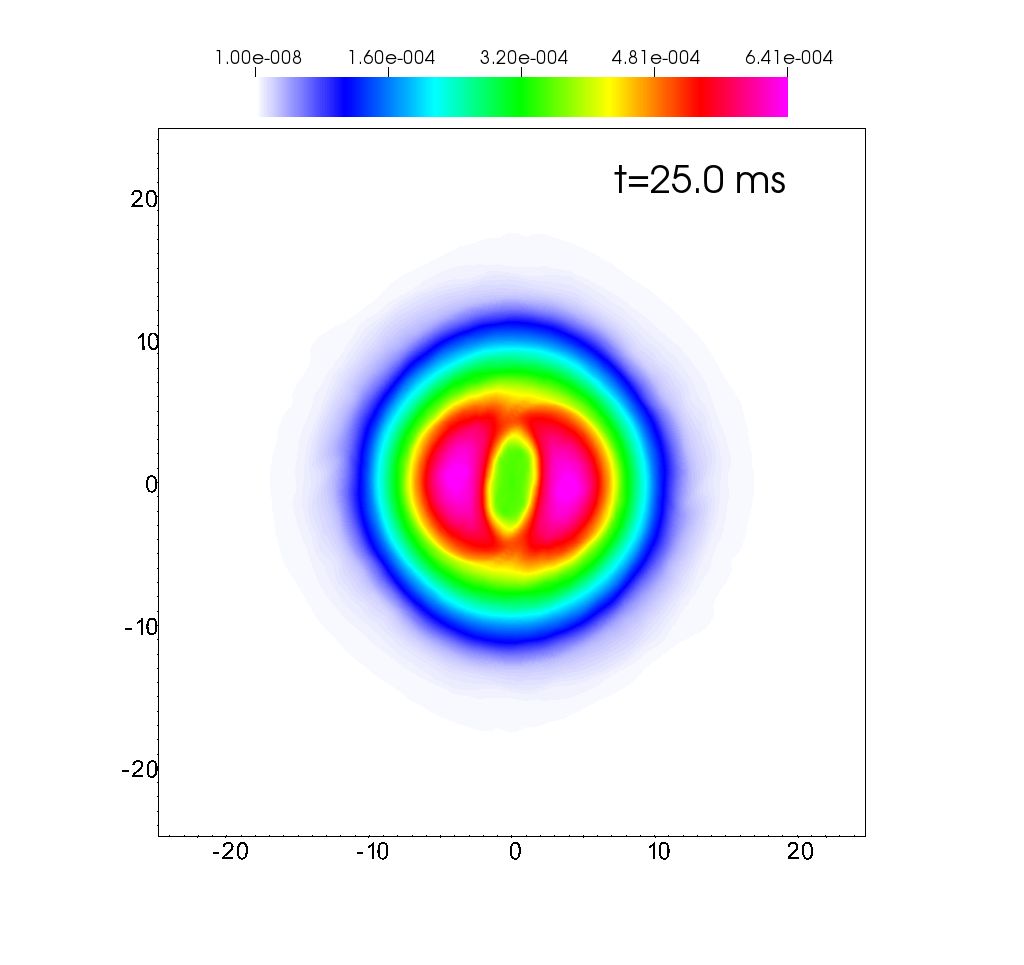}
 \includegraphics[width=0.45\textwidth]{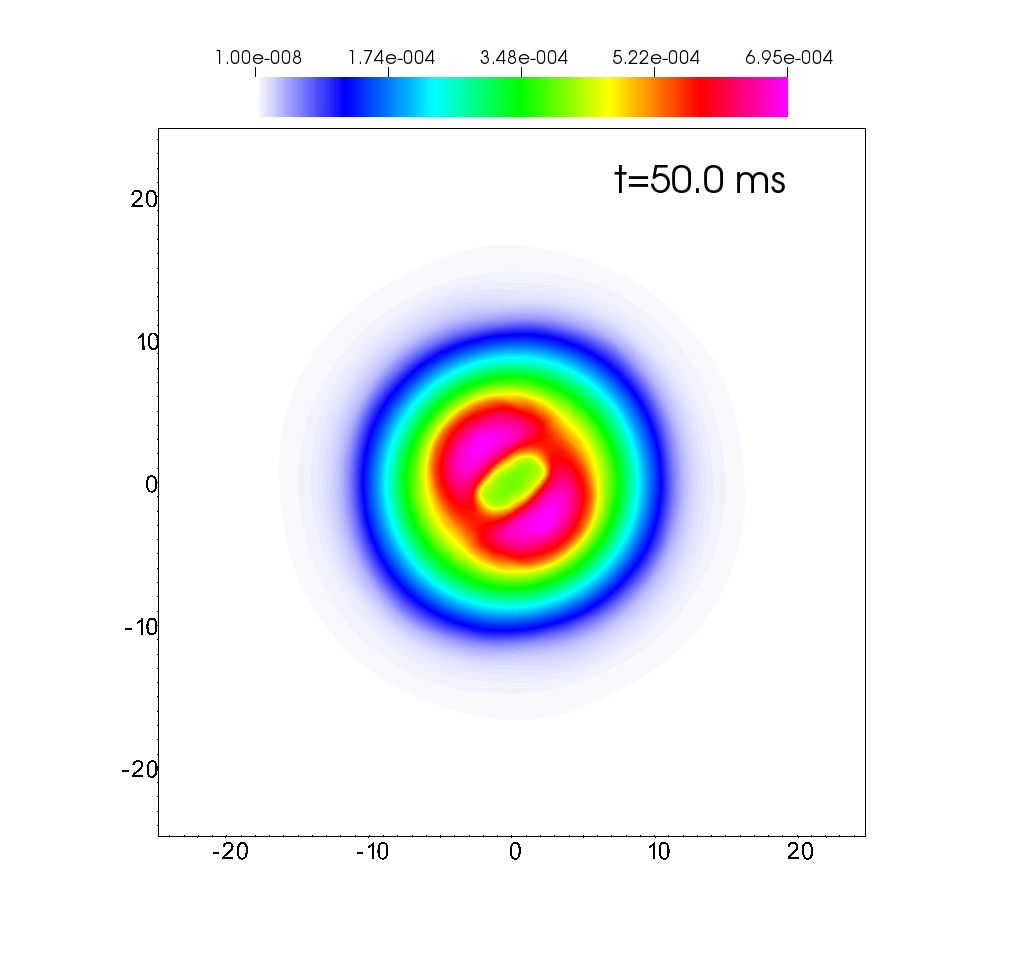}
 }
 \mbox{
 \includegraphics[width=0.45\textwidth]{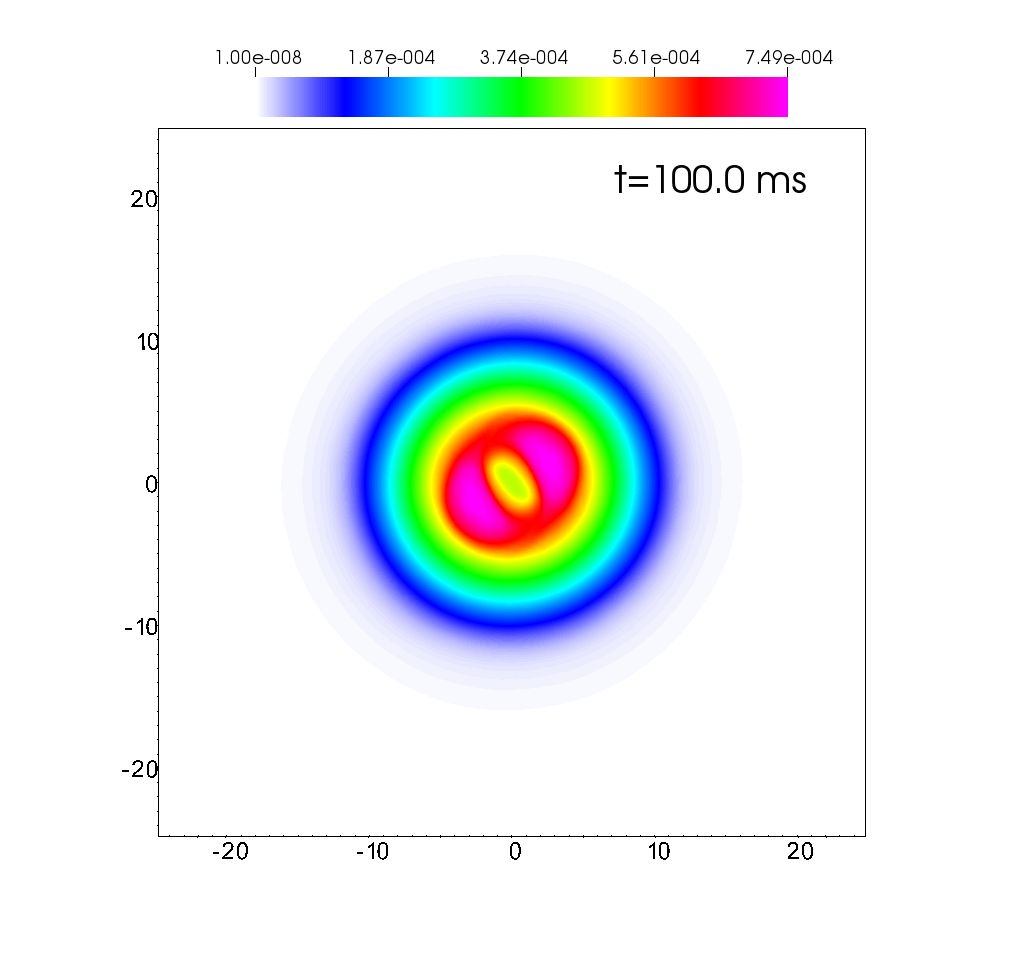}
 \includegraphics[width=0.45\textwidth]{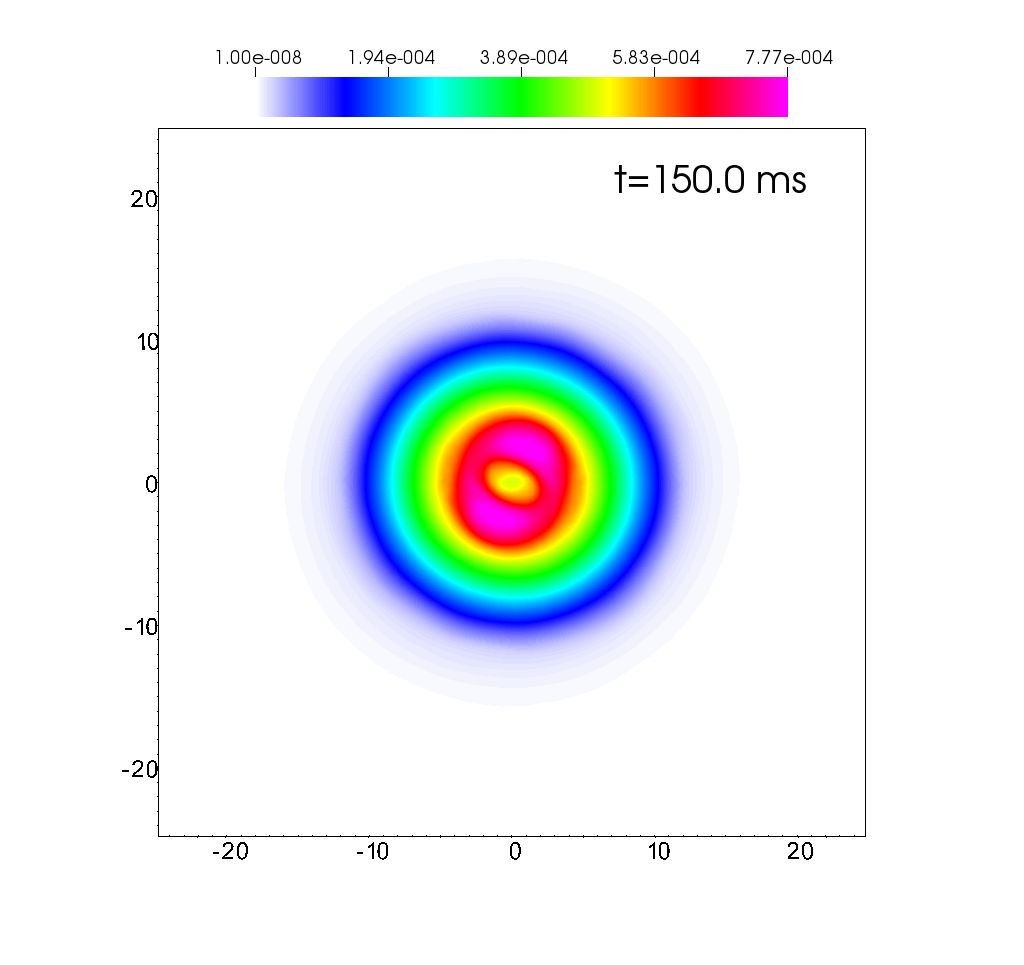}
 }

 \caption{\label{fig_rho_144_t} Snapshots of the rest-mass density $\rho$ in the equatorial $x$-$y$ plane at progressive simulation time. The series are made from a typical evolution with the initial NS baryon mass $M_b=1.445M_\odot$ using the high resolution $\rmd x=9/64=0.140625$. Color schemes range from $\rho_\mathrm{max}(t)$ to $10^{-8}$ $c^6/G^3M_\odot^2$ ($= 6.177\times 10^9 \mathrm{g}/\mathrm{cm}^3$) in each plot. Also see Fig. \ref{fig_rho_180} (the lower panel) for the snapshot taken at $t=180.0$ ms from the same evolution. }
\end{figure}

\begin{figure}[htbp!]
 \centering
 \mbox{
 \includegraphics[width=0.45\textwidth, trim={0, 100, 0, 0}, clip]{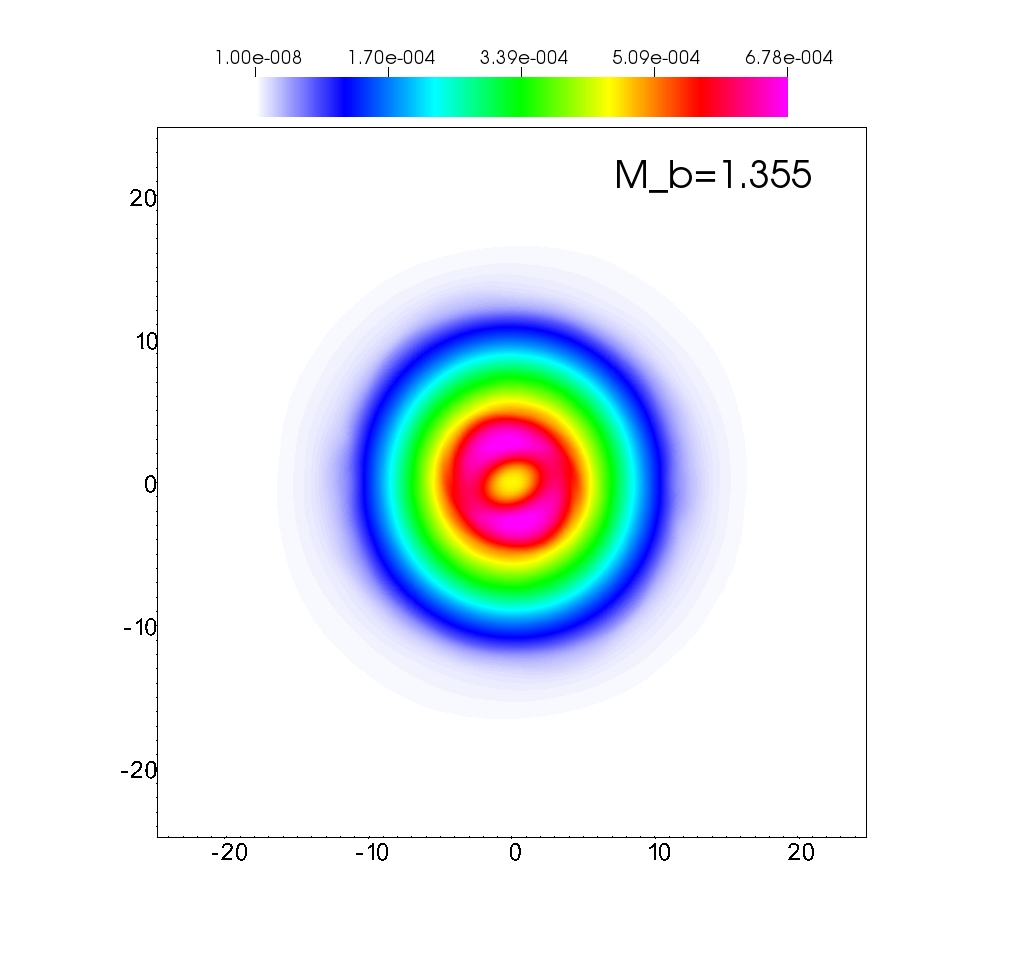}
 \includegraphics[width=0.45\textwidth, trim={0, 100, 0, 0}, clip]{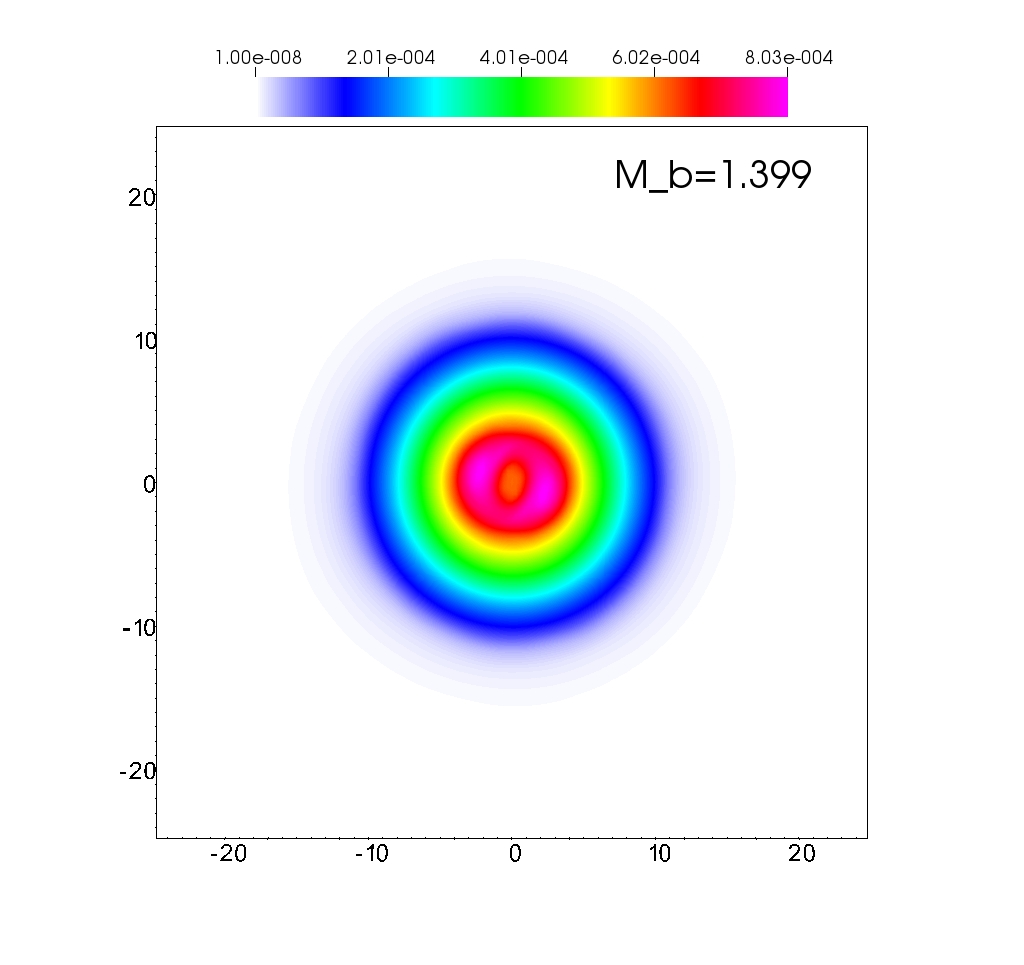}
 }
 \mbox{
 \includegraphics[width=0.45\textwidth, trim={35, 300, 35, 325}, clip]{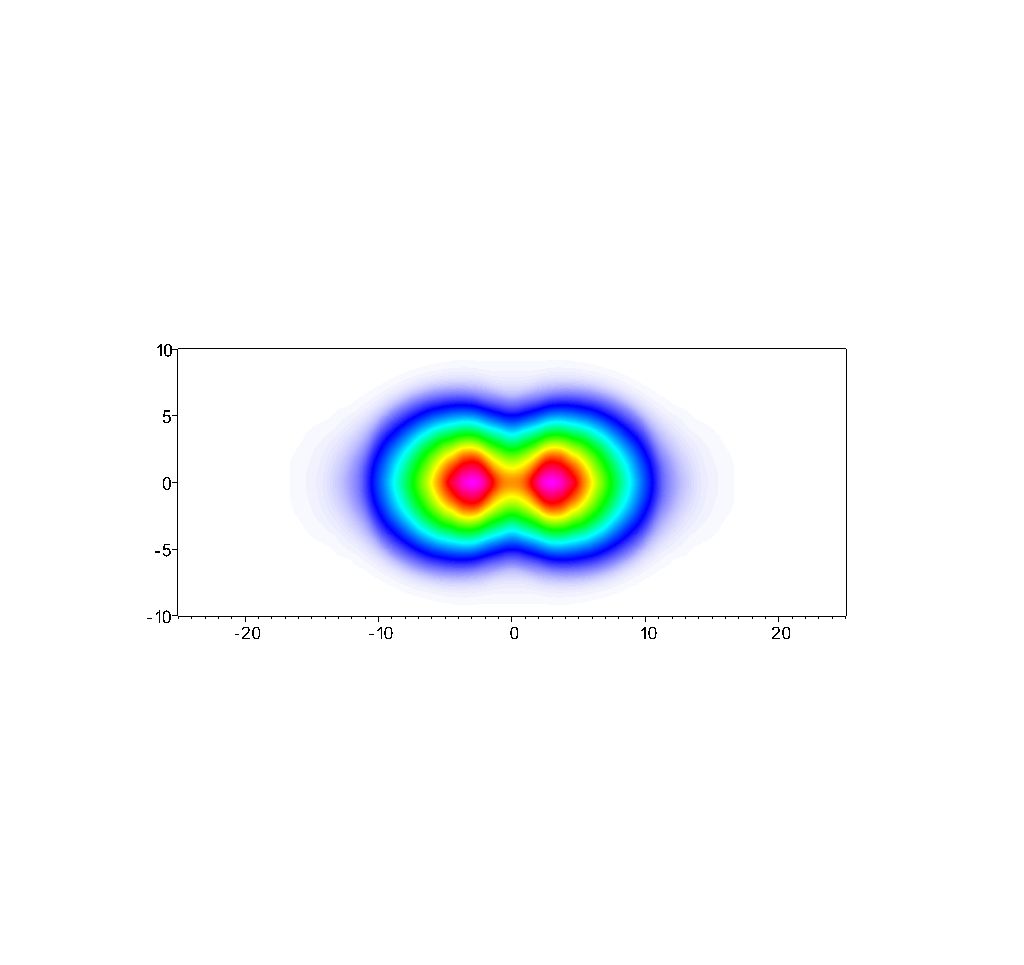}
 \includegraphics[width=0.45\textwidth, trim={35, 300, 35, 325}, clip]{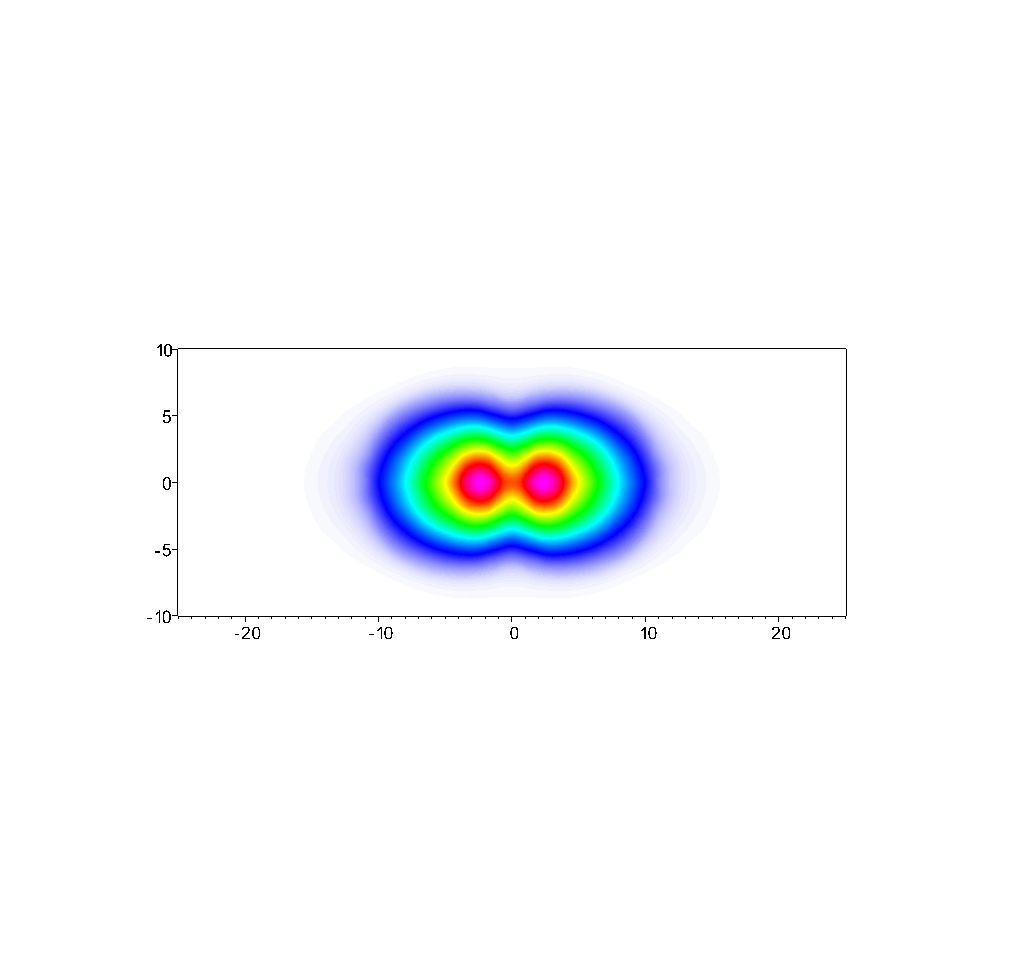}
 }
 \includegraphics[width=0.45\textwidth, trim={0, 100, 0, 0}, clip]{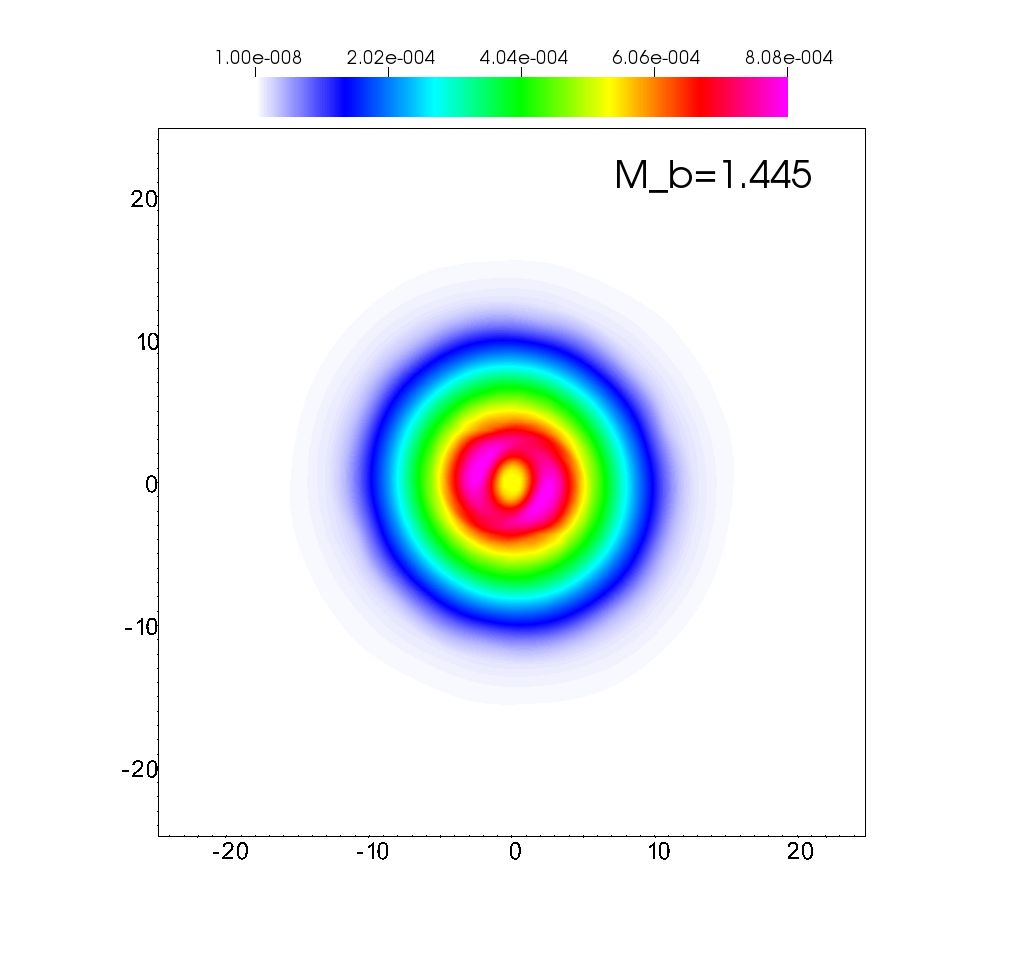} \\
 \includegraphics[width=0.45\textwidth, trim={35, 300, 35, 325}, clip]{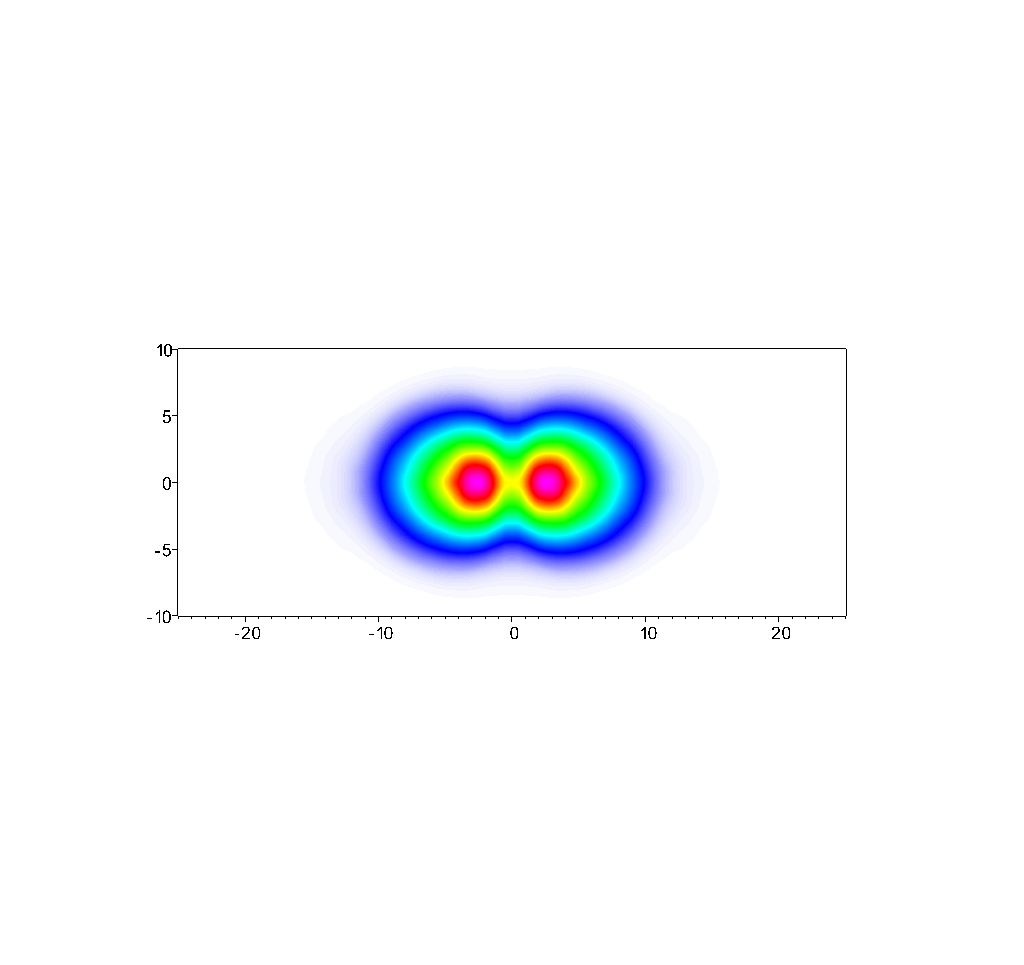}
 \caption{\label{fig_rho_180} Comparison of the rest-mass density snapshots in the equatorial $x$-$y$ plane and the meridian $x$-$z$ plane at the simulation ending time $t=180.0$ ms (i.e., about 170 ms after merger) for three different NS baryon masses $M_b=1.355$, $1.399$, $1.445 M_\odot$. Color schemes range from $\rho_\mathrm{max}(t)$ to $10^{-8}$. }
\end{figure}

\begin{figure}[htbp!]
 \centering
 \includegraphics[width=0.9\textwidth]{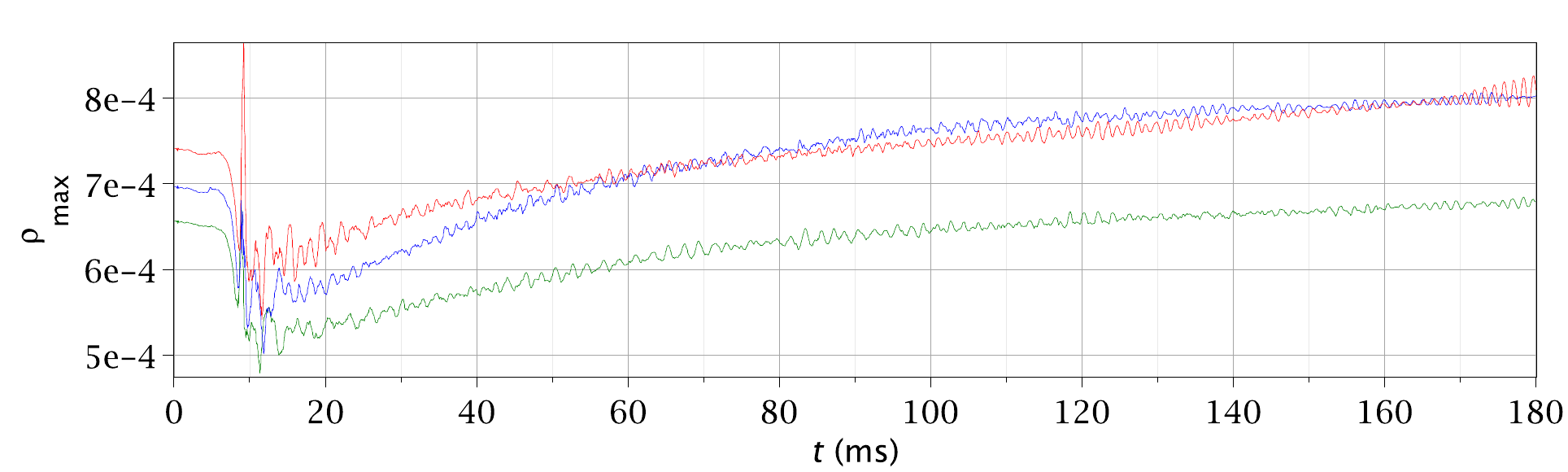}
 \caption{\label{fig_rho_max} Time evolution of the maximum rest-mass density $\rho_\mathrm{max}$ for the three binary models with $M_b=1.355$ (green), $1.399$ (blue), $1.445 M_\odot$ (red). Note the wiggles on top of the ascending curves. }
\end{figure}

The overall matter dynamics is demonstrated in Fig. \ref{fig_rho_144_t}, where we have plotted rest-mass density snapshots of the typical evolution with NS baryon mass $M_b=1.445M_\odot$. The evolutions with lower masses are qualitatively similar to Fig. \ref{fig_rho_144_t} in the postmerger phase. In Fig. \ref{fig_rho_180} we compare the final density snapshots from the three runs when they are stopped at $t=180$ ms. All three postmerger processes are characterized by the formation of torus-like remnants with varying sizes and maximum rest-mass densities.

Pursuant to Fig. \ref{fig_rho_144_t}, the binary make tidal contact and begin to merge after about $t=8.3$ ms, or about 2.7 orbits, from the start of the simulation. Following a turbulent and violent merger phase (cores merging and bouncing), the system produces a double-core structure at about 15 ms with two separate high-density regions shaped in clearly defined crescents (half moon; marked red and purple). The concave sides of the crescents are facing each other, their arms enclosing an oval area of lower rest-mass density in between. From this point on, the postmerger evolution proceeds in a quasi-stationary fashion, since the GW dissipation timescale ($>200$ ms, \cite{Paschalidis17}) is much longer than the dynamical timescale. Coursing through the last four panels of Fig. \ref{fig_rho_144_t} from 25 ms to 150 ms, the double-crescent core spreads out in azimuthal directions, and gradually smears out to a torus. The overall shape of the core becomes rounder and reduces in size due to angular momentum loss through gravitational radiation. During this secular process, the hypermassive remnant oscillates in a regular and quasi-periodic manner, and is supported by large differential rotation against gravitational collapse.

The final outcomes of the three simulations terminated at $t=180$ ms are presented in Fig. \ref{fig_rho_180}. Instead of forming an ellipsoid as has been reported in many previous publications, the remnants take on a shape of thickened biconcave disks: depressed in the middle, with a dumbbell-like $x$-$z$ cross section, and a torus-shaped inner core surrounded by extended lower-density envelope. The maximum rest-mass density is not situated at the center, and its value increases for greater $M_b$ among the three plots (see the color schemes). We point out that at $t=180$ ms the high density area (marked purple) is still not evenly spread along the ring (note slight bumps). It indicates the matter distribution not being quite axisymmetric. But we do anticipate that a higher degree of axisymmetry (but oscillating) will be attained after $t=200$ ms (see Sec. \ref{sec:resltn}).

As mentioned in the introduction, differentially rotating NS models in axisymmetric toroidal shapes have been built with presumed rotational profiles. They are widely tested for dynamical instability of isolated NSs. In our work, such models have been consistently generated from BNS mergers, hence improving from previous artificial models. We also mention that (proto-)NSs with toroidal cores have also been reported in fully GR simulations of stellar-core collapse with sufficiently differentially rotating progenitors \cite{Shibata05b}, as well as dynamical-capture BNS mergers \cite{Paschalidis15}. Therefore our results should not be a total surprise.

To understand the rapidly rotating toroidal configuration, we comment that its formation requires the relaxation timescales of matter redistribution to be smaller along circular directions than along radial directions. This is more likely to occur in stiff EOSs which tend to have less compact and more uniform density profiles. For nuclear EOSs that are not so stiff as the $\Gamma=2$ model, the formation of toroidal cores, if it ever occurs as a transient state, may not be so prominent as the ones presented here.

The maximum rest-mass density drifting away from the center is a consequence of a large amount of residual angular momentum after merger. The fast differential rotation provides centrifugal support that pushes matter outwards and creates a centrally loose configuration. Therefore it is not possible for the remnant stars in our simulations to collapse in the toroidal shapes, which can only occur after the tori contract and the cores become compact enough. Hereby we can propose a dynamical mechanism that toroidal core formation can substantially delay collapse into BHs and prolong the lifetime of hypermassive metastable remnants from merger. This is in contrast with dumbbell/bar-shaped cores.

As mentioned before, the NS masses of our models are all well below the threshold value ($2M_b\sim 3.4 M_\odot$) for prompt collapse. The mass difference among them is less than $7\%$, and hence similarity in their postmerger evolutions is anticipated. The fact that toroidal remnants show up in all three models reflects a level of robustness of our results as well as qualitative consistency at the adopted resolution. But irregularities also exist. Note that the torus size of the medium-mass run ($M_b=1.399M_\odot$) appears smaller than the other two in Fig. \ref{fig_rho_180}. The corresponding maximum density also rises above the one in the high-mass run ($M_b=1.445M_\odot$) after $t=80$ ms in Fig. \ref{fig_rho_max}. They suggest possible underlying complications of long-term simulations.

Fast spinning NSs with toroidal density profiles are subject to the dynamical bar-mode ($m=2$) instability \cite{Tohline90}, which is universal regardless of EOSs. The bar deformation induces ellipticity and generates gravitational radiation. Hence the annular remnants are secularly unstable, and sustained GW emission alters the structure of differentially rotating remnants. Strong rotation will also incur non-axisymmetric f-mode/r-mode oscillations due to GW-driven instability. We will come back to this point in the next subsection.


\subsection{GW spectra}

\begin{figure}[htbp!]
 \centering
 \includegraphics[width=0.9\textwidth]{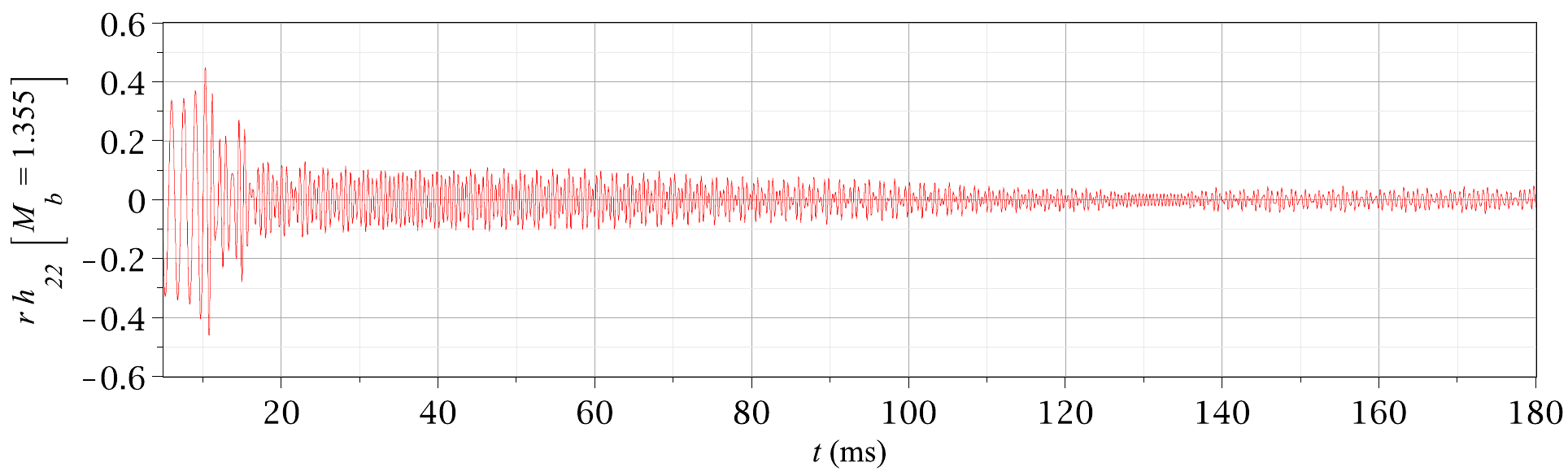}
 \includegraphics[width=0.9\textwidth]{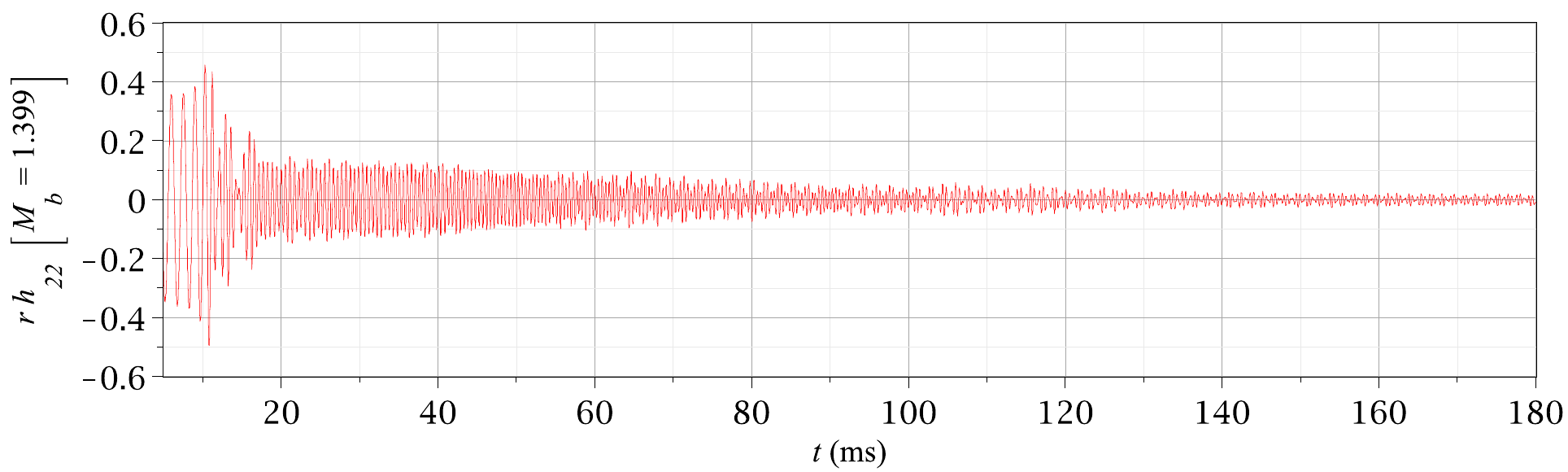}
 \includegraphics[width=0.9\textwidth]{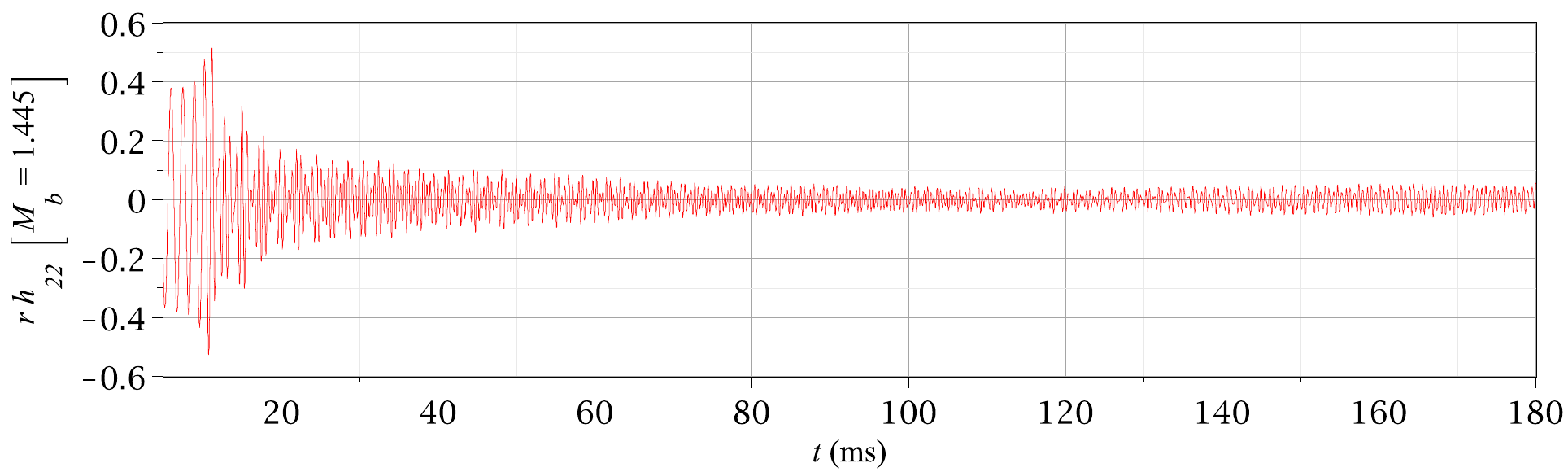}
 \caption{\label{fig_strain} Gravitational strain $r h_{22}^+$ from the three binary models with NS baryon masses $M_b=1.355$, $1.399$, $1.445 M_\odot$ and evolved using the high resolution with the finest grid size $\rmd x=9/64=0.140625$. The signals are extracted at the radius $r=500$, and $t$ denotes the coordinate time. The plots are prepared in high resolution. }
\end{figure}

\begin{figure}[htbp!]
 \centering
 \includegraphics[width=0.9\textwidth]{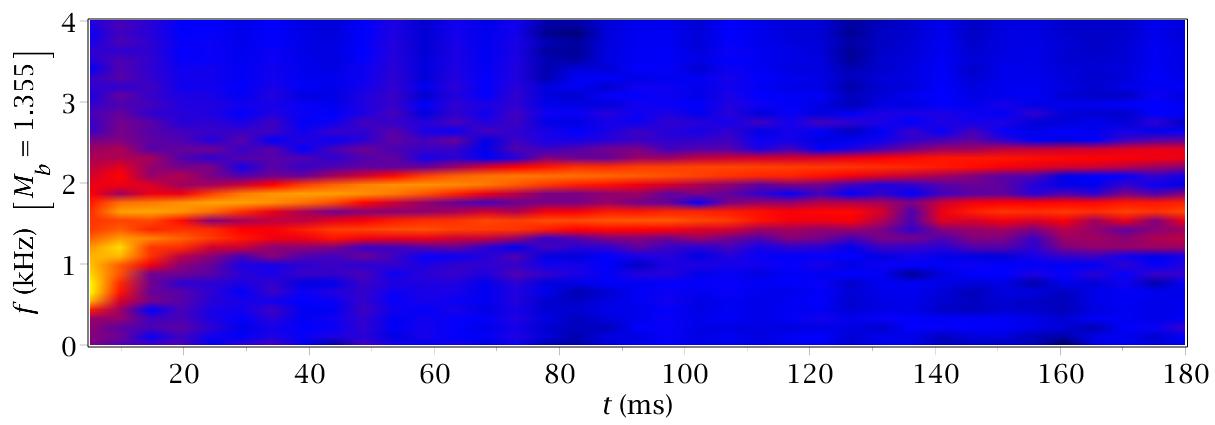}
 \includegraphics[width=0.9\textwidth]{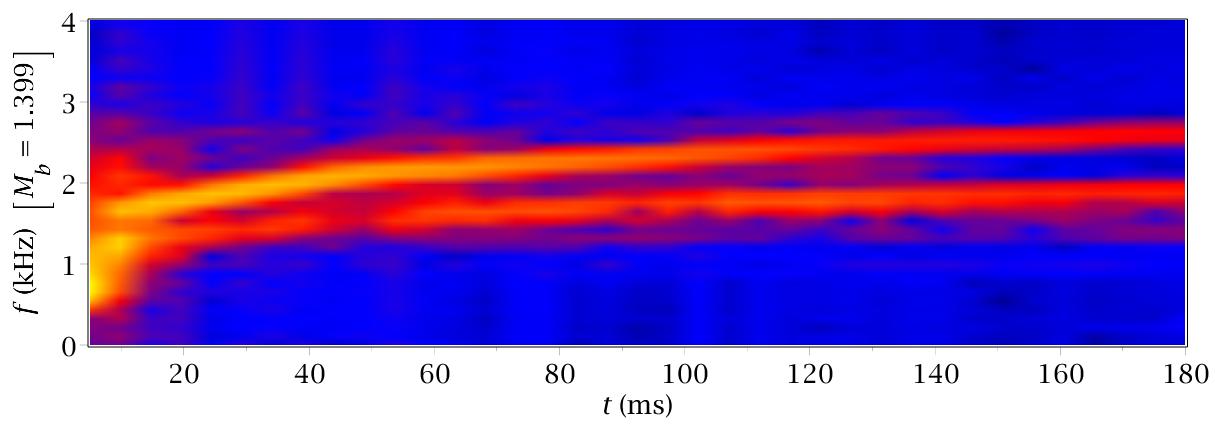}
 \includegraphics[width=0.9\textwidth]{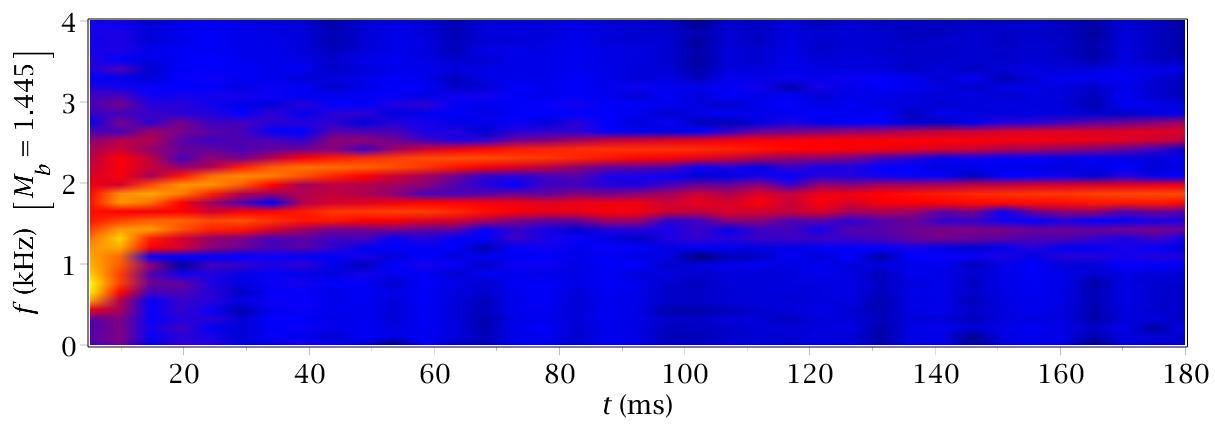}
 \caption{\label{fig_spec} Spectrograms of the GW strain $r h_{22}^+$ in Fig. \ref{fig_strain}. Three peaks are identifiable after the coordinate time $t\sim 100$ ms, denoted by $f_2$, $f_{2-0}$, and $f_C$ respectively from top to bottom. For the PSD plot, see Fig. \ref{fig_PSD} and check Table \ref{tab_freq} for calculated peak frequencies. }
\end{figure}

Figs. \ref{fig_strain} and \ref{fig_spec} present the long waveforms, i.e., the dominant $l=2=m$ component of $h_+$, and the corresponding spectrograms, organized from top to bottom with increasing NS masses. The signals are extracted at the radius $r=500$. We cut out the first 5 ms that contains spikes caused by the idealized initial data. The plots in Fig. \ref{fig_strain} are prepared in high resolution and allow zooming in for finer details.

In Fig. \ref{fig_strain}, the three strain signals attain their highest amplitudes at about $t=11$ ms after inspirals, then enter into a turbulent phase up to $t=17$ ms during which the amplitudes change sporadically. Afterward, the strains gradually decay till $t \sim 110$ ms, and then sustain fairly steady strength. There is even slight increase in the amplitudes of the low and high mass plots (the first and third panels of Fig. \ref{fig_strain}) after 140 ms. One might have originally expected that the strain amplitude should decrease monotonically (e.g., exponential decay). However this is not the case here. We can explain this by temporary growth and saturation of GW-driven unstable oscillation modes which induce additional GW radiation.

As usual, the strain frequencies increase dramatically following merger. For the most part of the postmerger waveforms, one is able to discern a modulation of the quasi-periodic signals from as early as $t=20$ ms. It indicates that the systems evolve in a quasi-stationary manner and that the postmerger GW spectra should contain multiple peaks. These are evidenced in Fig. \ref{fig_spec}. Now we summarize most prominent long-term postmerger spectral properties in the following.

(1) The three spectrograms are qualitatively similar to one another in regard of general patterns and tendencies of the dominant spectral lines. The lines steadily rise up as the NS compactness increases due to energy loss through GW radiation, which results in faster spinning. Also note that the overall position of the spectral lines tends to shift upwards for higher mass binaries as they are more compact.

(2) Two clearly recognizable dominant frequencies persist throughout the entire postmerger simulation. This is in contrast with single dominant frequencies reported in previous literature (e.g., \cite{Takami15,Rezzolla16}). To help understanding this major difference, we comment that the double peaks are keenly associated with the double cores inside the remnants \cite{Shibata03} that also oscillate quasi-radially with a non-negligible amplitude. In published simulations that used softer nuclear EOSs, the double-core structure only shows up very briefly (e.g., \cite{Shibata05a,Hanauske16}) and then fuse into an ellipsoid which is significantly more compact and structurally simpler. In accordance, the sub-dominant peaks that have existed initially diminish. Therefore the difference can be traced down to greater compactness generically associated with softer EOSs.

(3) We denote the frequency of the higher dominant spectral line by $f_2$ (we ignore the overtone $f_3$ since it is short-lived \cite{Takami15,Rezzolla16}). The physical origin of $f_2$ is clear and it corresponds to the quadrupole $l=2=m$ mode. The values of $f_2$ can be measured from the data. They are equal to twice the rotational frequencies of the double cores or, at later stages, the bar-mode deformation of the toroidal cores. The $f_2$ frequency has bee shown to correlate with the tidal coupling constant \cite{Bernuzzi15}. Its value increases over time but the rate of increase slows down as GW emission levels off. The notation here follows \cite{Takami15,Rezzolla16} and is dubbed $f_\mathrm{peak}$ in \cite{Bauswein12a,Bauswein12b,Bauswein15}.

\begin{figure}[htbp!]
 \centering
 \includegraphics[width=0.9\textwidth]{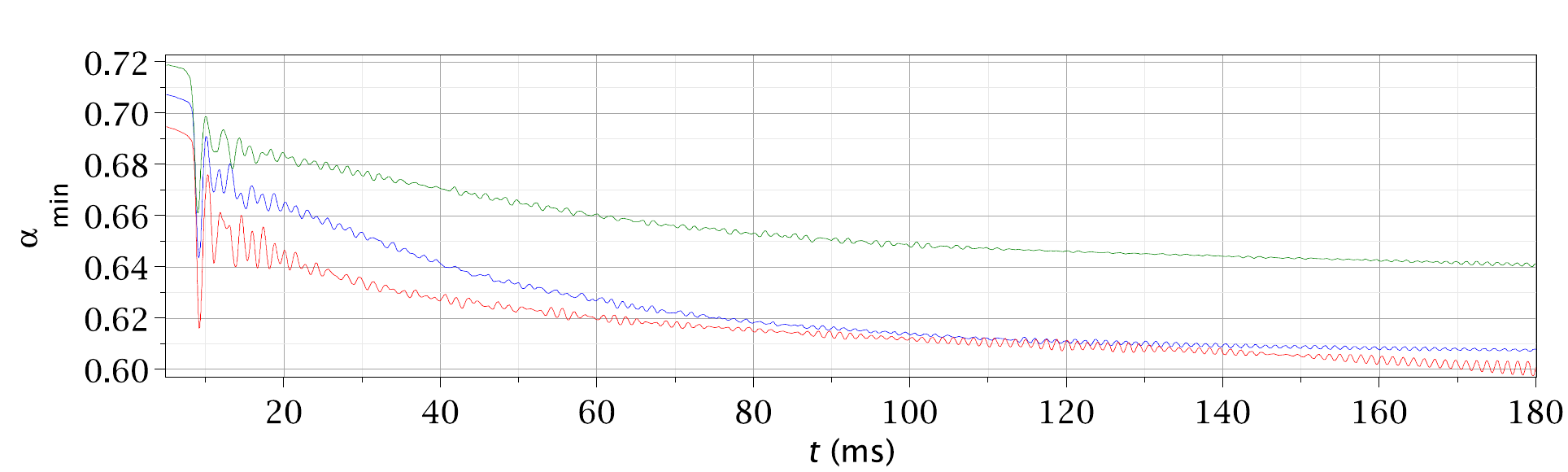}
 \caption{\label{fig_alpha} Time evolution of the minimum of the lapse function $\alpha$ for the three binary models with $M_b=1.355$ (green), $1.399$ (blue), $1.445 M_\odot$ (red). Note the wiggles on top of the descending curves. }
\end{figure}

(4) We denote the frequency of the lower dominant spectral line by $f_{2-0}$, following \cite{Stergioulas11}. As the name suggests, it represents a mode coupling between quadrupole modes and quasi-radial oscillation modes ($m=0$) \cite{Shibata03,Stergioulas11}. The latter can be pictured as the double cores undergoing compression and expansion inside the gravitational potential well, initially excited by radial plunging in the late inspiral phase. Its frequency can be measured by tracking the undulations in the minimum values of the lapse functions (Fig. \ref{fig_alpha}) or the maximum values of the rest-mass density. Both quantities reflect the NS compactness. We denote the quasi-radial oscillation frequency by $f_0$. By measuring $f_0$, $f_2$, and $f_{2-0}$ separately, we find them following $f_{2-0} \simeq f_2-f_0$ (see Table. \ref{tab_freq}).

\begin{figure}[htbp!]
 \centering
 \includegraphics[width=0.8\textwidth]{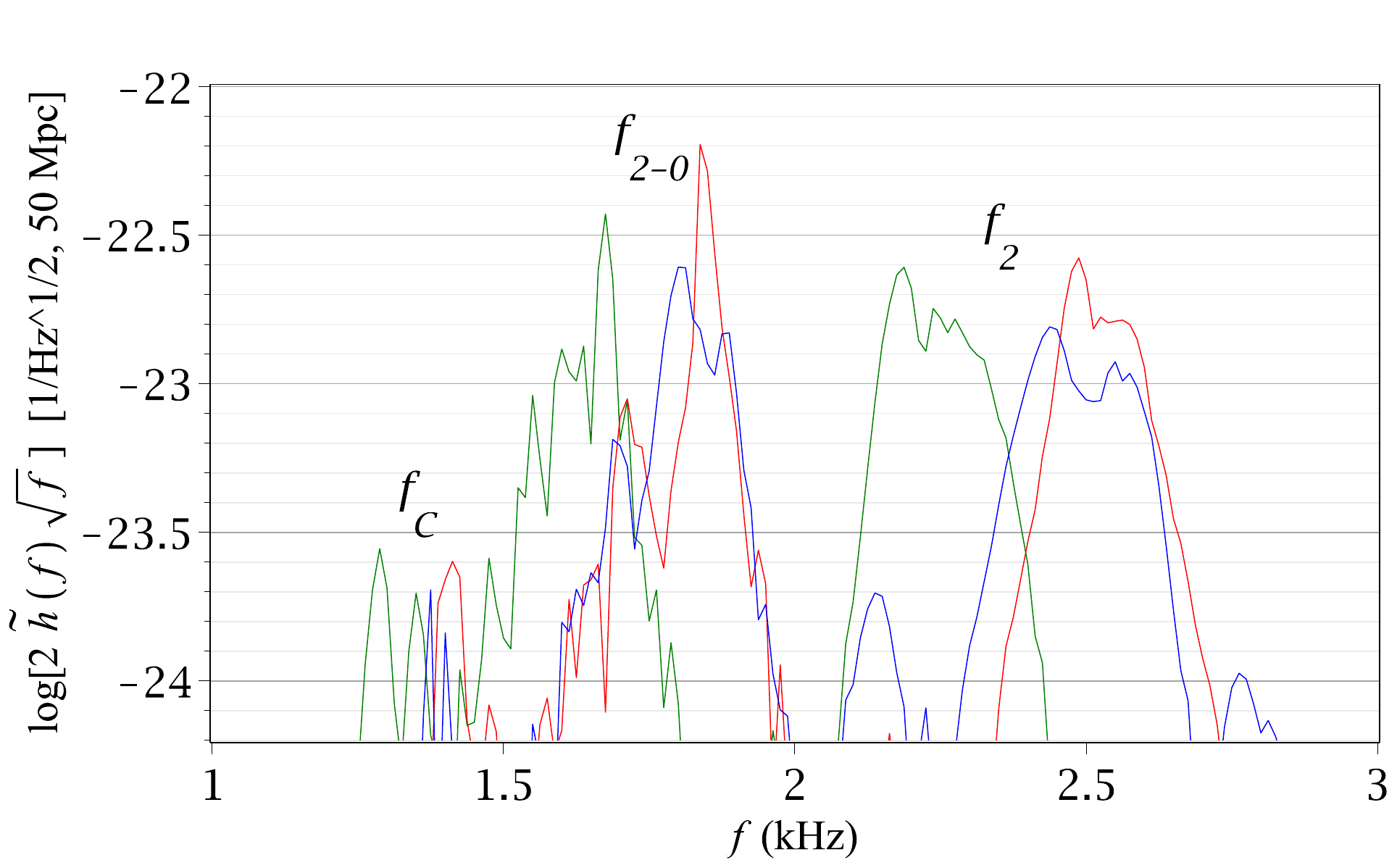}
 \caption{\label{fig_PSD}  PSD over the 80 ms window $t\in[100-180]$ ms for the three binary models with $M_b=1.355$ (green), $1.399$ (blue), $1.445 M_\odot$ (red). The sources are at a distance of 50 Mpc. The intensity of $f_C$ is expected to continue to grow after $t=180$ ms. }
\end{figure}

\begin{table}
 \caption{\label{tab_freq} List of the frequency peaks $f_0$, $f_2$, $f_{2-0}$, and $f_C$ (kHz) calculated from the PSD over the 30 ms window $t\in[150-180]$ ms. }
 \begin{indented}
 \lineup
 \item[]\begin{tabular}{cccccc}
 \br
   Model               & $f_2$ & $f_0$ & $f_2-f_0$ & $f_{2-0}$ & $f_C$  \cr
 \mr
   $M_b=1.355 M_\odot$ & 2.33  & 0.78  & 1.55      & 1.67      & 1.30   \cr
   $M_b=1.399 M_\odot$ & 2.60  & 0.78  & 1.82      & 1.90      & 1.40   \cr
   $M_b=1.445 M_\odot$ & 2.60  & 0.74  & 1.86      & 1.87      & 1.43   \cr
 \br
 \end{tabular}
 \end{indented}
\end{table}

To explain the strength of $f_{2-0}$, we comment that the quasi-radial oscillations may correlate with motions of the antipodal bulges (ellipticity) in the outer layer that rotate slower than the double cores (more obvious in animation). It is analogous to the rotating spiral-shaped deformation $f_\mathrm{spiral}$ introduced in \cite{Bauswein15}. We suspect that the frequency may overlap $f_{2-0}$ and contribute the peak as well.

(5) At about 100 ms, one can clearly see the emergence of a sub-dominant third peak $f_C$ positioned lower than the two dominant peaks in the spectrograms (even as early as 60 ms in the medium-mass case $M_b=1.399 M_\odot$). This is confirmed by the PSD plot in Fig. \ref{fig_PSD}, and also presented in simulations with lower resolutions (see the next section). The peak relates to the remnant oscillation as the stars evolve into a new phase. The exact nature of this new peak deserves more future investigation. Our speculation would be unstable oscillations (f-mode/r-mode) driven by the secular Chandrasekhar-Friedman-Schutz (CFS) instability \cite{Chandrasekhar70,Friedman78a,Friedman78b}. These unstable modes have long been proposed in isolated rotating NSs, which may lead to detectable GW signals (see, e.g., the review articles \cite{Andersson03,Stergioulas03,Paschalidis16,Andersson11} and references therein). Recently the issue has also been reexamined in the context of BNS merger (e.g., \cite{Doneva15}). To our best knowledge, this is the first time that evidence of their excitation is reported in fully general-relativistic BNS simulations that last long enough to reach their onset. It validates BNS mergers as an astrophysical scenario that can excite unstable oscillation modes. The processes also indicate that the non-axisymmetric imprint of BNS mergers slowly fades away as the systems enter into a new dynamical phase taken over by different mechanisms. The GW signals that contain new peaks may convey unique signatures of the NS structure.

(6) As mentioned in the previous subsection, we point out that there appears to be some abnormalities in the medium-mass run with $M_b=1.399 M_\odot$. As Fig. \ref{fig_strain} shows, it emits stronger gravitational radiation during 20-50 ms than the other two, which explains the smaller torus in Fig. \ref{fig_rho_180}, and hence the weaker emission after 140 ms. Moreover, the spectrum does not seem as ``clean'' as the other two. We are uncertain about the underlying cause.


\section{Long-Term Effect of Resolution} \label{sec:resltn}

\begin{figure}[htbp!]
 \centering
 \includegraphics[width=0.8\textwidth]{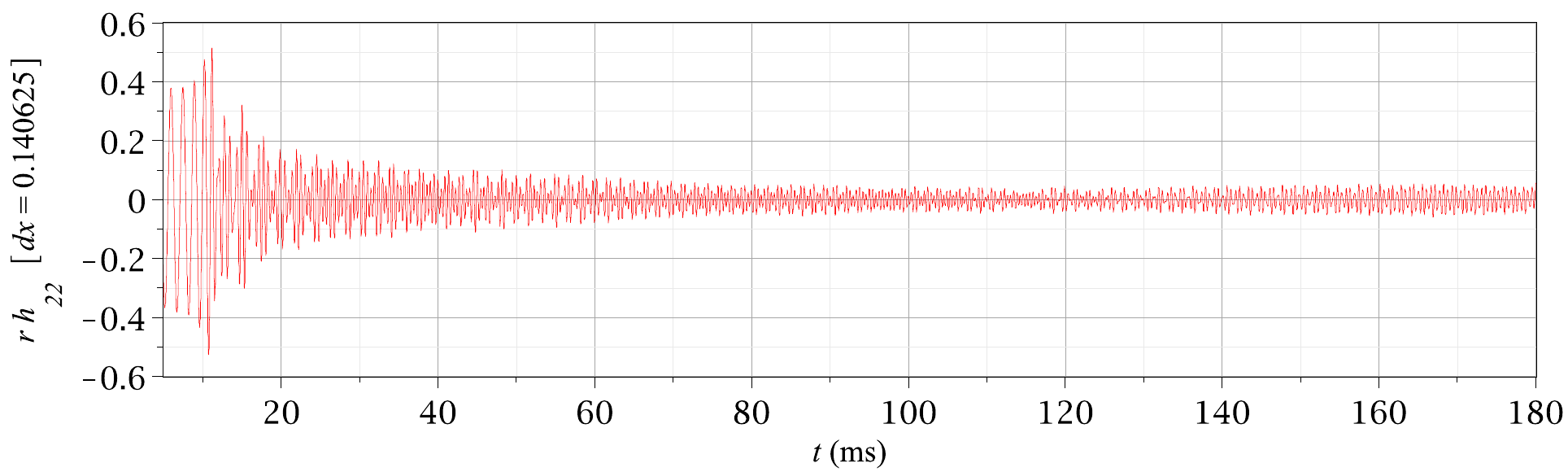}
 \includegraphics[width=0.8\textwidth]{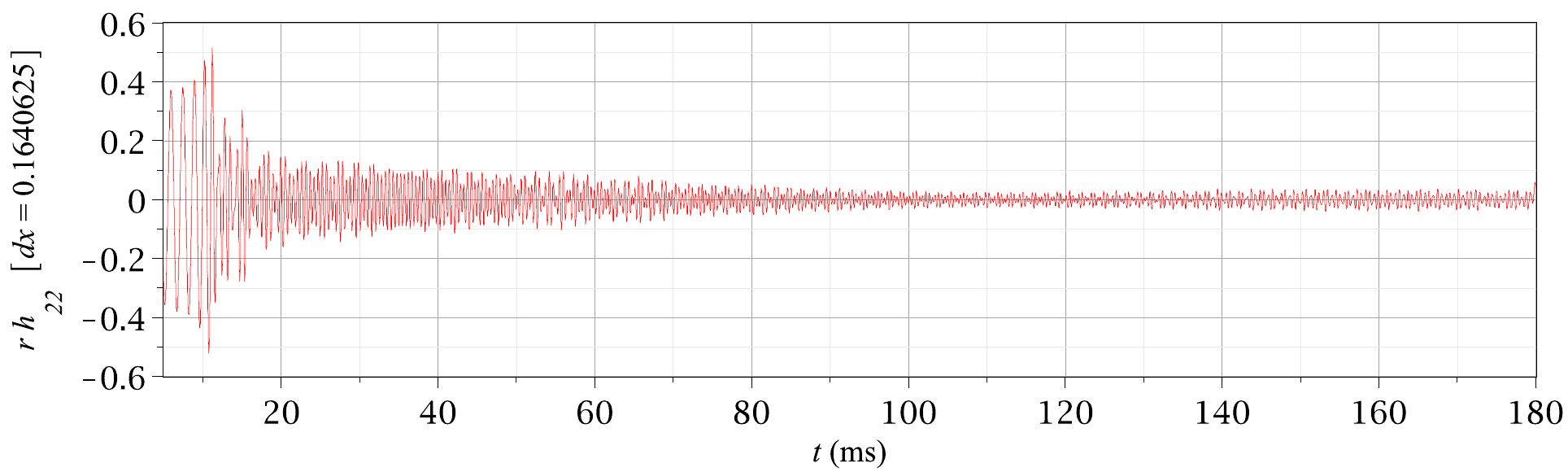}
 \includegraphics[width=0.8\textwidth]{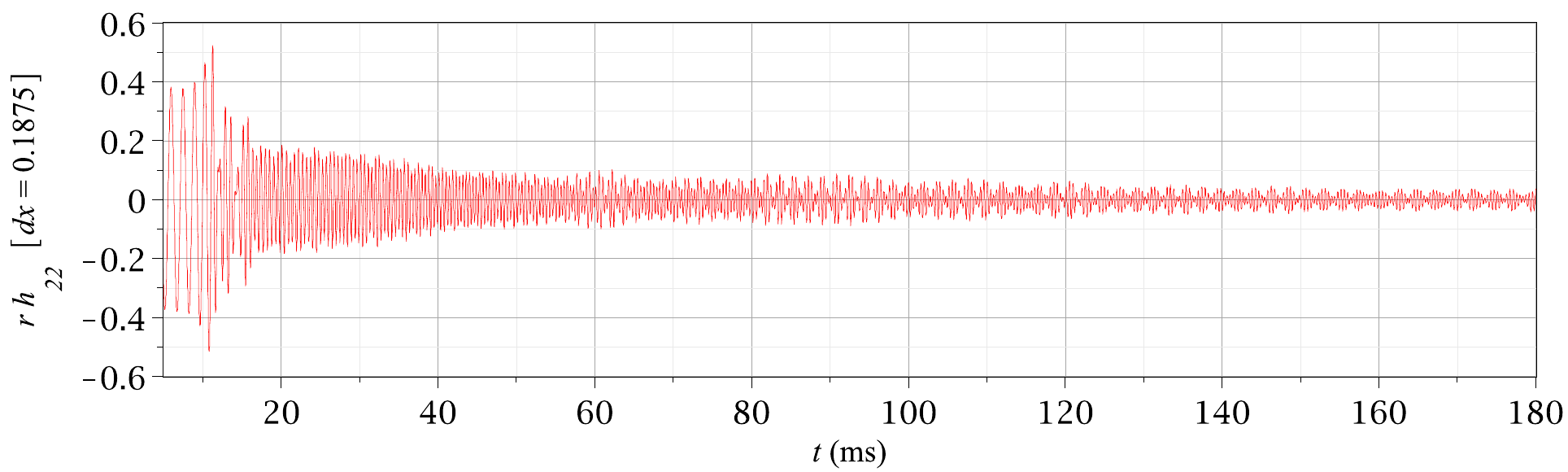}
 \includegraphics[width=0.8\textwidth]{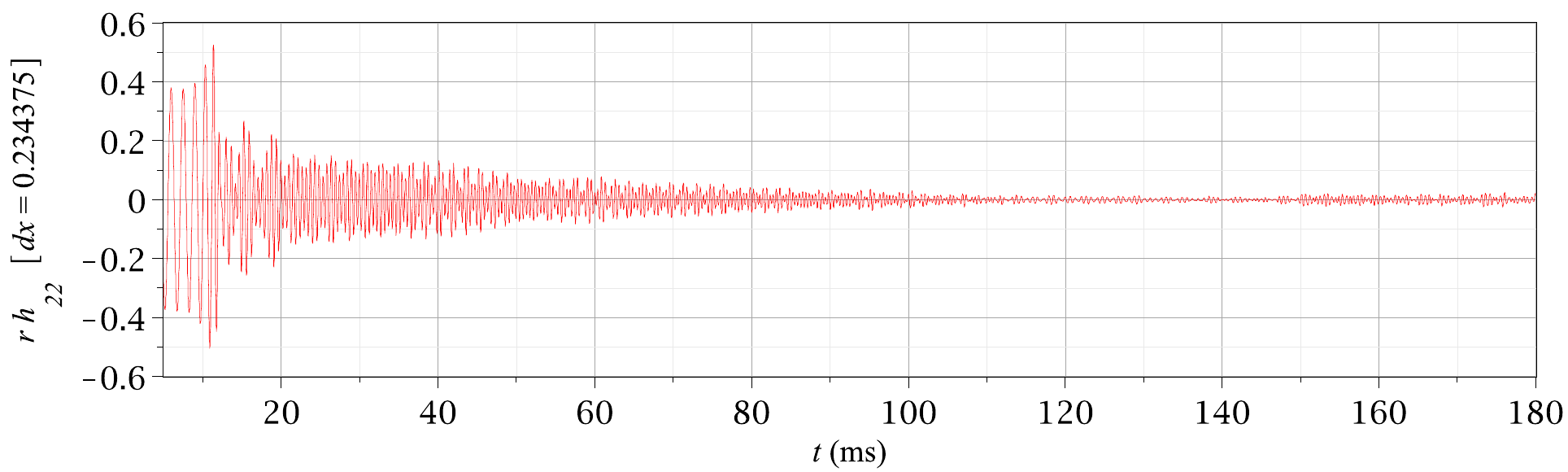}
 \includegraphics[width=0.8\textwidth]{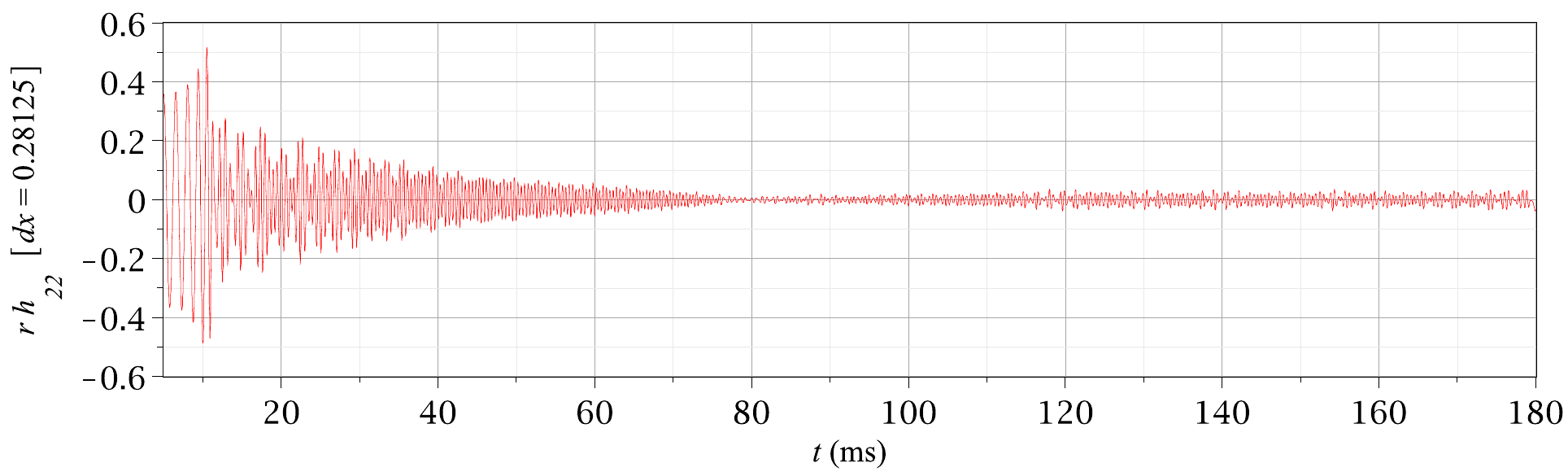}
 \caption{\label{fig_144_strain} Gravitational strain $r h_{22}^+$ from the binary models with NS baryon mass $M_b=1.445 M_\odot$ and evolved using five resolutions (zoom in for finer details). The finest grid sizes are $\rmd x=9/64$, $10.5/64$, $12/64$, $15/64$, $18/64$ ($=0.28125$) from top to bottom. The signals are extracted at the radius $r=500$, and $t$ denotes the coordinate time. }
\end{figure}

\begin{figure}[htbp!]
 \centering
 \includegraphics[width=0.8\textwidth]{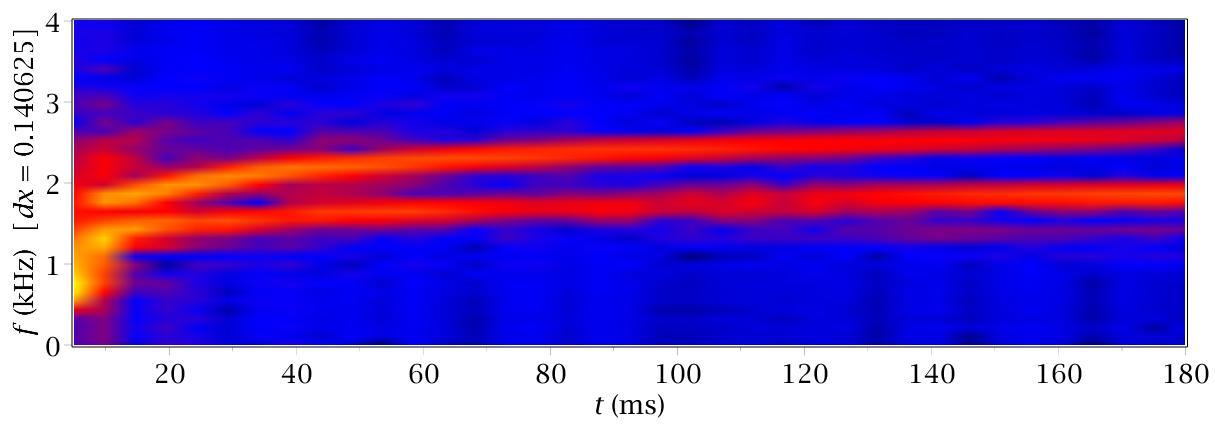}
 \includegraphics[width=0.8\textwidth]{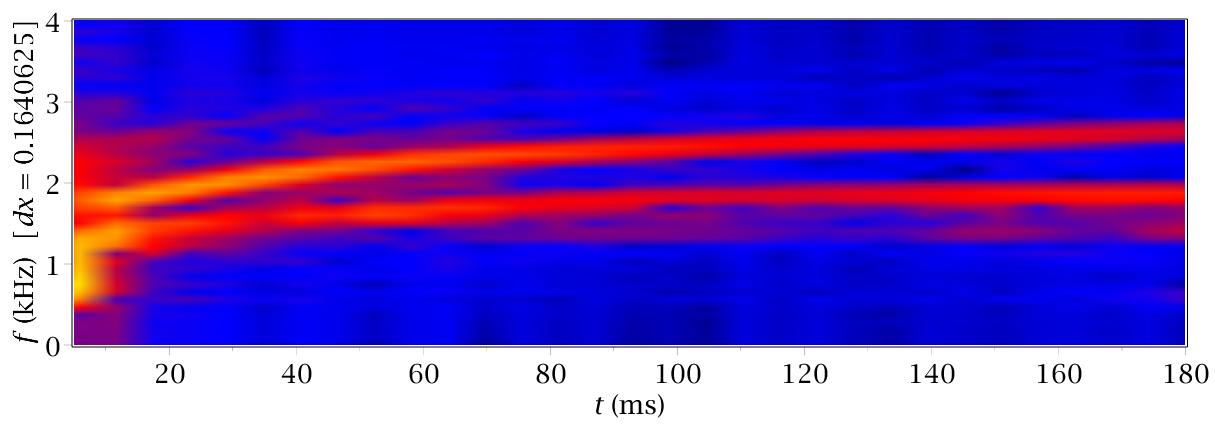}
 \includegraphics[width=0.8\textwidth]{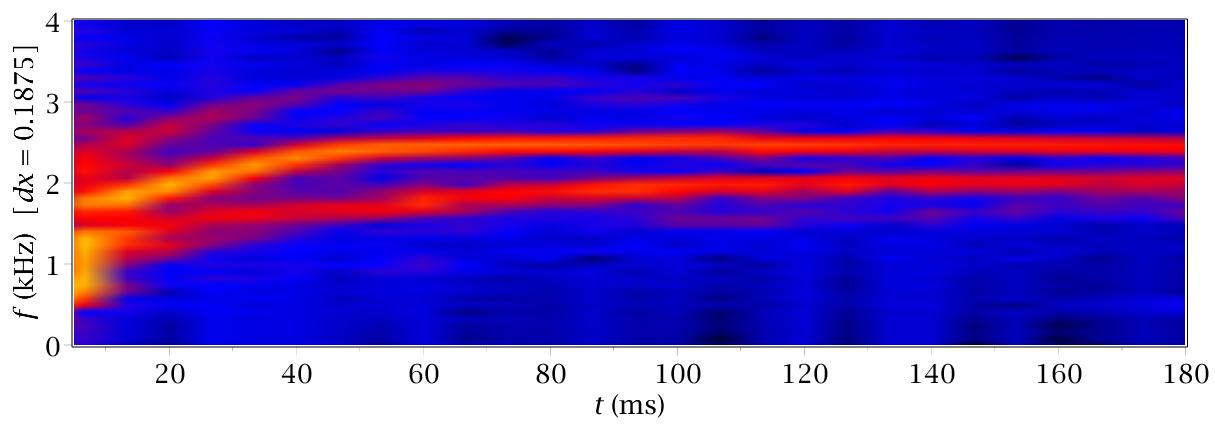}
 \includegraphics[width=0.8\textwidth]{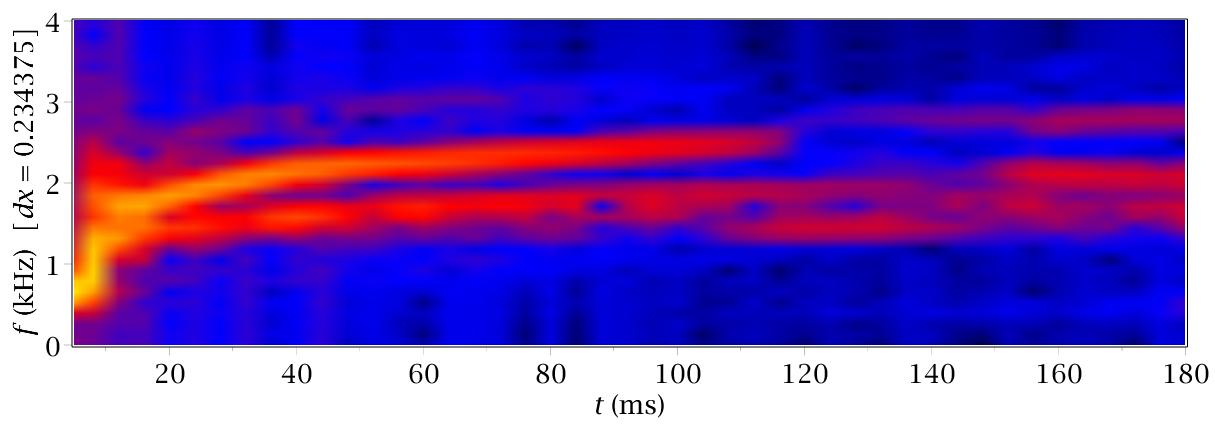}
 \includegraphics[width=0.8\textwidth]{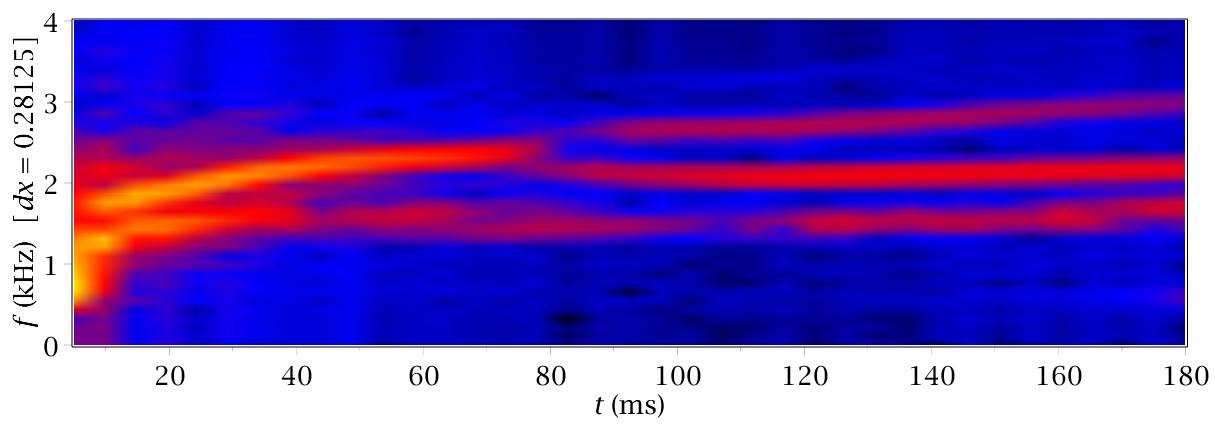}
 \caption{\label{fig_144_spec} Spectrograms of the GW strain $r h_{22}^+$ in Fig. \ref{fig_144_strain}.}
\end{figure}

\begin{figure}[htbp!]
 \centering
 \includegraphics[width=0.4\textwidth]{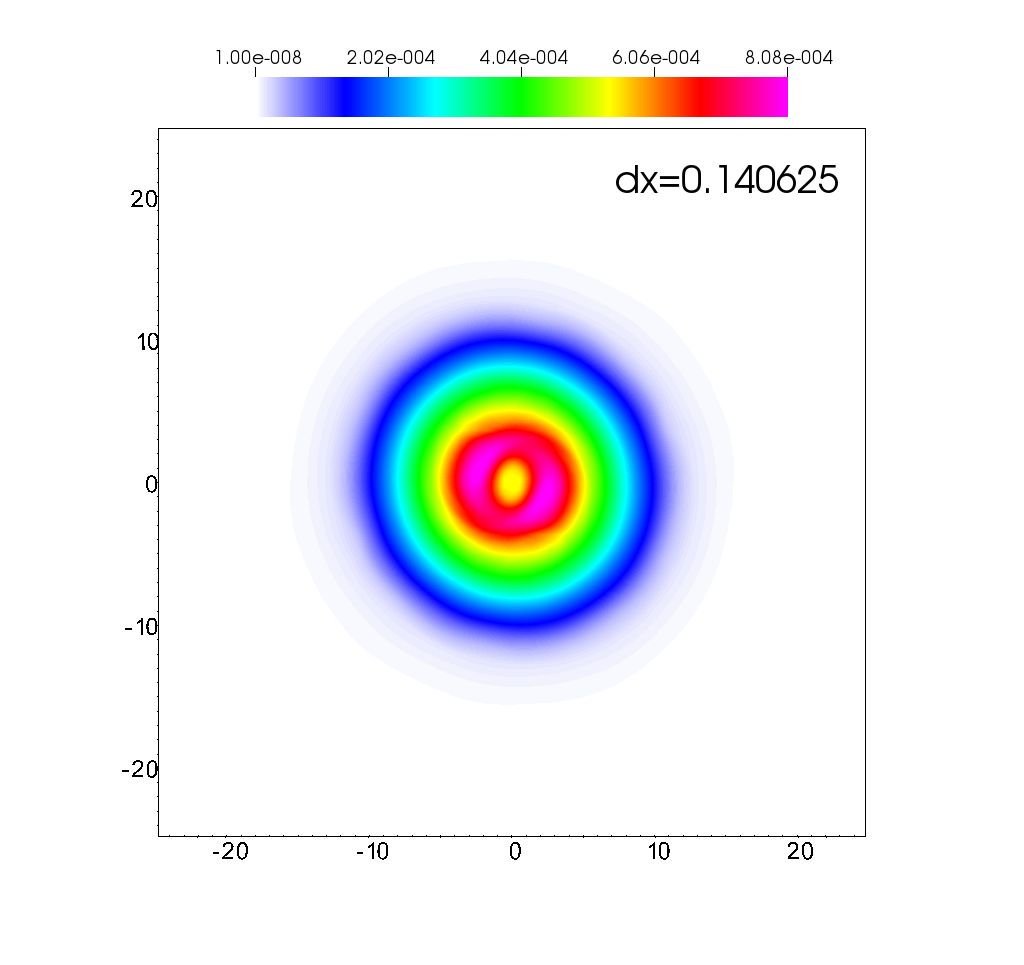}
 \mbox{
 \includegraphics[width=0.4\textwidth]{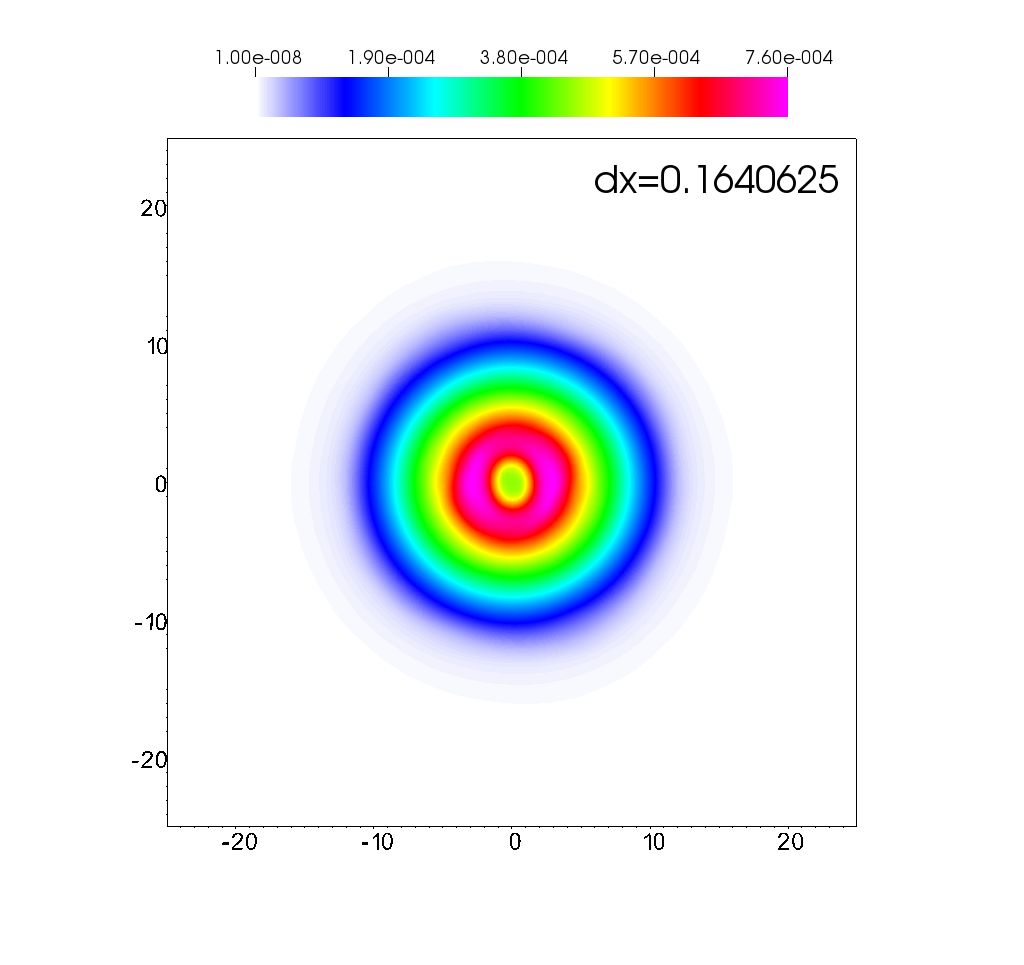}
 \includegraphics[width=0.4\textwidth]{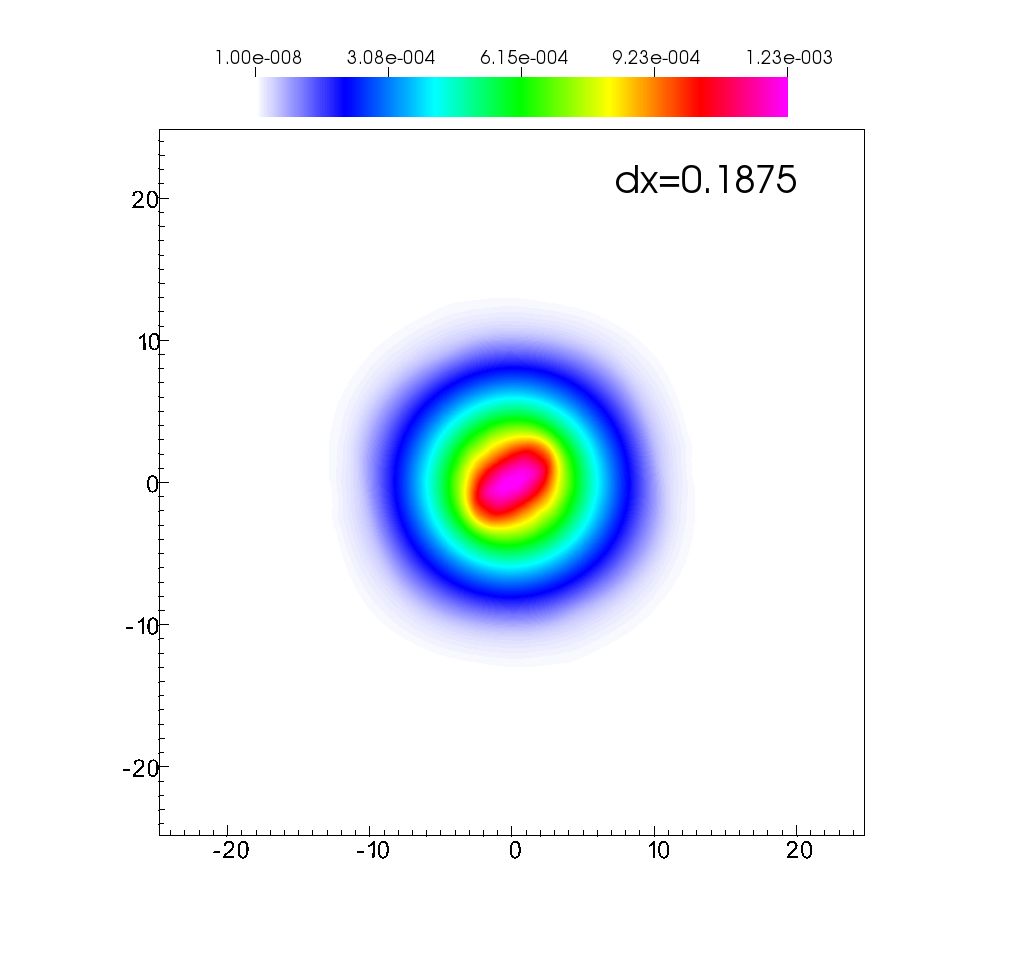}
 }
 \mbox{
 \includegraphics[width=0.4\textwidth]{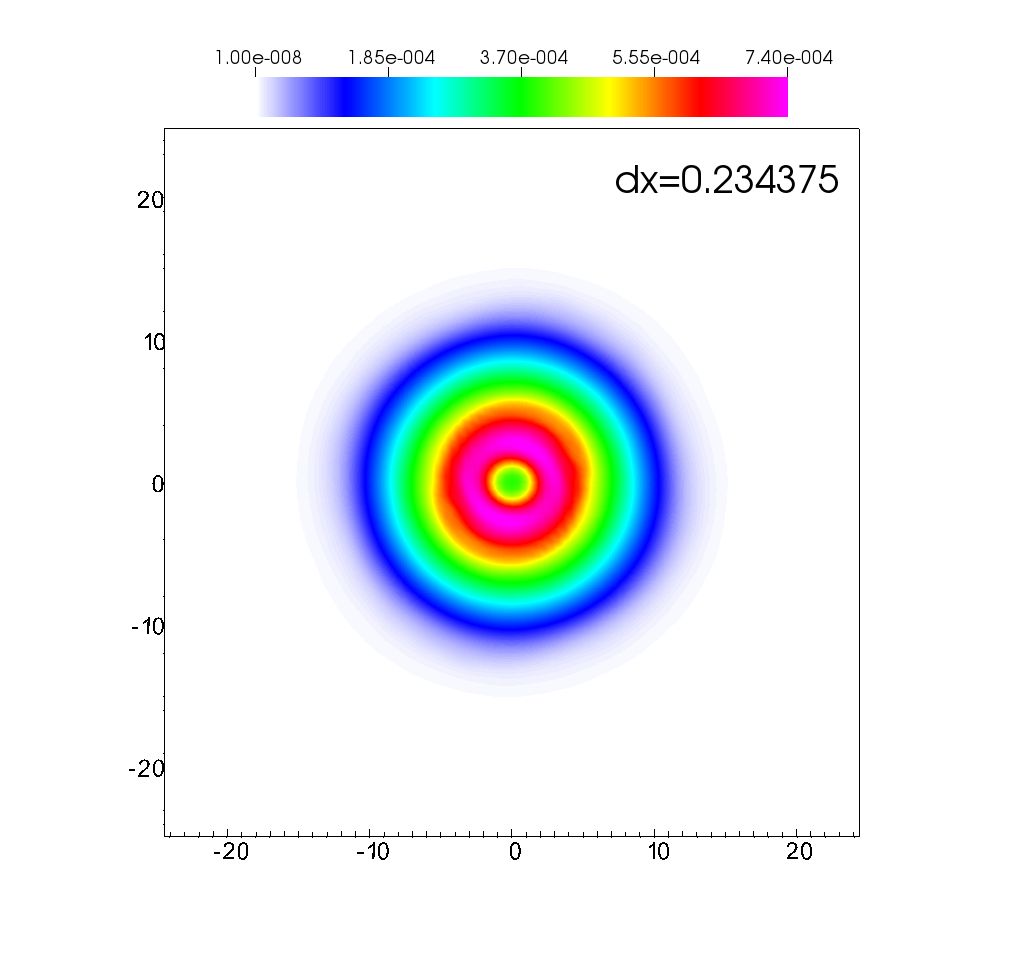}
 \includegraphics[width=0.4\textwidth]{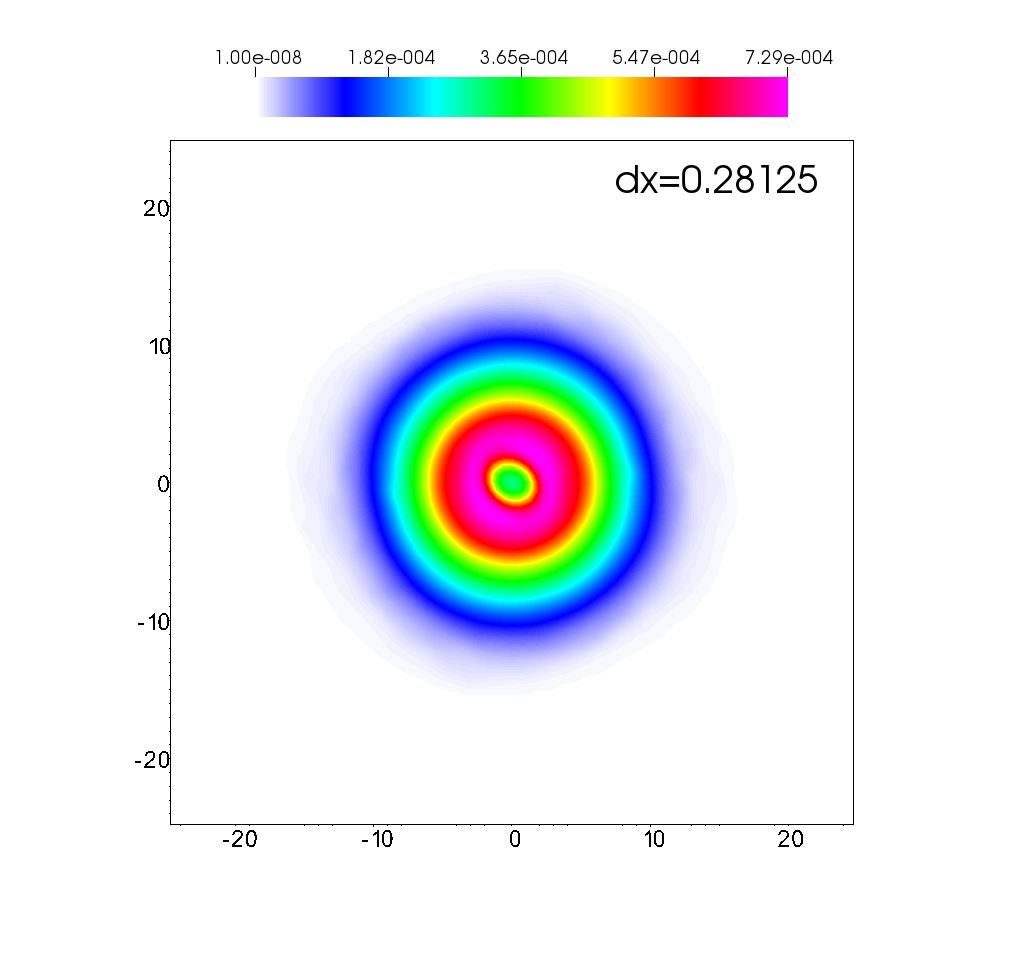}
 }
 \caption{\label{fig_144_rho} Comparison of the rest-mass density snapshots in the equatorial $x$-$y$ plane at the simulation ending time $t=180.0$ ms (i.e., 170 ms after the merger) using five grid resolutions $\rmd x=9/64$, $10.5/64$, $12/64$, $15/64$, $18/64$ ($=0.28125$). Color schemes range from $\rho_\mathrm{max}(t)$ to $10^{-8}$. }
\end{figure}

In this section, we concentrate on the initial data model $M_b=1.445 M_\odot$ and study the qualitative effect of grid resolution on the GW spectral properties and remnant dynamics. Five spatial resolutions are considered which have the finest grid sizes from $\rmd x=9/64=0.140625$ to $\rmd x=18/64=0.28125$. The chosen range has been made wider than those normally tested for numerical convergence. As mentioned before, it is impractical to expect waveform convergence over long-term simulations with current numerical schemes and available computational resources. Far more important in this work is to capture reliable qualitative behavior. As far as grid resolution is concerned, we are more interested in the qualitative changes and trends that different resolutions bring about, as well as in checking the consistency of our results (see, e.g., \cite{Takami15,Rezzolla13}). Therefore in the test, we will be focusing on two major aspects, i.e., GW spectral patterns and remnant morphology. Indeed, over the extended test range, the postmerger simulation of 170 ms shows nontrivial dependence on the grid resolution.

Fig. \ref{fig_144_strain} summarizes the waveforms from the five simulations, organized from top to bottom with increasing grid sizes. The mergers (highest amplitudes) all occur at about 11 ms. Fig. \ref{fig_144_spec} shows the corresponding spectrograms. Fig. \ref{fig_144_rho} collects the rest-mass density snapshots at the end of the simulations $t=180$ ms. The plots with $\rmd x=9/64$ are duplicates from the previous section, and we put them here for ease of comparison. Some interesting qualitative features are the following.

(1) The two highest resolution runs ($\rmd x=9/64$, $10.5/64$) show good overall agreement with each other. Significant phase difference only begins to show after 20 ms. Though the strain amplitude in $\rmd x=10.5/64$ is systematically lower than $\rmd x=9/64$ after $\sim 100$ ms, their spectral patterns are similar and they both produce ring-shaped remnants at $t=180$ ms, with less compactness and a rounder shape in $\rmd x=10.5/64$. Emergence of a third low-frequency peak $f_C$ is quite discernible in both spectrograms. The consistency adds extra confidence to our results in the previous section.

(2) As the grid size increases, one expects that the inherent numerical dissipation becomes more significant in affecting postmerger simulations, since numerical viscosity scales with grid spacing. The effect appears rather evident in the two lowest resolution runs ($\rmd x=15/64$, $18/64$) which share some common features. Their strain amplitudes are both characterized by a brief quenching at $\sim 140$ ms in $\rmd x=15/64$ and at $\sim 80$ ms in $\rmd x=18/64$. This is in accordance with the remnants settling into a nearly axisymmetric metastable intermediate state in the shape of a torus, which results in diminished quadrupole radiation. The short ``pause'' in GW emission does not last long before non-axisymmetric unstable oscillation modes kick in and reach plateau afterwards (e.g., note the oval green inner circle in Fig. \ref{fig_144_rho}). In the spectrograms, the transition is clearly reflected in breakage of spectral lines, where the initial spectral pattern that contains non-asymmetric imprint of BNS merger is taken over by unstable oscillation modes of rotating NSs. It naturally divides the postmerger evolution into two phases that exhibit differing characteristics.

If one takes into account that numerical viscosity alters the timescale of the remnant dynamics, the behavior in the two low-resolution runs is thus indicative to the high-resolution runs ($\rmd x=9/64$) beyond 180 ms. Hence we predict that a similar transition of the GW spectral pattern may also take place in the high-resolution simulations after $t=200$ ms, as the mass distribution of the remnants turns more axisymmetric. Meanwhile the $f_C$ peak is likely to reach a higher intensity. Nonetheless the entire transition may not be so clean-cut as in $\rmd x=18/64$ (cf. Fig. \ref{fig_144_spec}).

The situation starting from $t\sim 140$ ms in $\rmd x=15/64$ and $t\sim 80$ ms in $\rmd x=18/64$ resembles the simulations carried out for determining instability of rotating NSs (e.g., \cite{DePietri14,Loffler15}). In those studies, one first constructs rotating NS initial data in axisymmetric stationary equilibrium (uniform rotation or the $j$-constant rotation law), and then evolves the models to determine the subsequent evolution. The advantage in our case is that such ``initial'' data models have been self-consistently generated from BNS merger simulations, even though they are affected by grid resolution. But we recall that the numerical code regains higher orders of convergence in the quasi-stationary stage. Hence by assuming these intermediate states as initial data, the subsequent simulation is still trustworthy even with relatively low resolution. By comparing our results with the previous studies, we find that the three-peak spectral structure in the two low-resolution runs resembles greatly to the ones in \cite{Zink10}, suggesting that the peaks may have an origin of unstable f-modes.

(3) Compared with the four simulations discussed above, the middle resolution run with $\rmd x=12/64$ appears rather puzzling in that a bar-shaped remnant is formed at $t=180$ ms instead (Fig. \ref{fig_144_rho}). We have confirmed this result by running the simulation for the second time. There is no qualitative difference and the ellipsoid remains. The exact underlying cause is unclear. We tentatively attribute this discrepancy to the turbulent dynamics during the merger phase (a few ms after the merger) when the convergence rate of the numerical schemes drops significantly \cite{Baiotti08,Baiotti09}. The subsequent evolution may sensitively depend on the merger phase. The influence of numerical errors may be amplified by the extended simulation time.

(4) The comparison indeed raises a caveat for long-term simulations. A check on consistency should be performed to ensure qualitative soundness of the results. There can be a least required grid resolution before one entering into a regime of qualitative consistency in regard to GW spectral patterns and remnant morphology, which is yet weaker than the requirement of full numerical convergence. Despite qualitative differences in certain aspects, all five simulations can reproduce the low-frequency peak $f_C$, suggesting a type of robust behavior independent of grid resolution.


\section{Conclusions} \label{sec:conclu}

The field of NR hydrodynamic simulation nowadays has matured significantly to provide decent accuracy and stability so that long-term BNS simulations may be attempted. It is hoped that one will obtain more realistic configurations of isolated fast rotating NSs from extended postmerger simulations. As a first step towards tracking secular evolution of long-lived BNS remnants, we have presented new results on postmerger dynamics over a 170 ms timescale and in simplified physical settings. The quasi-stationary remnants have evolved for long enough so that new phenomena begin to manifest. The simulation time is not limited by numerical stability, which has proven remarkable in the {\tt Einstein Toolkit}, but more likely by numerical dissipation. Although we have used a simple $\Gamma=2$ ideal-fluid EOS, the work can help to set up a possible range of dynamical behavior that more realistic EOSs may give rise to. We summarize the main conclusions and remarks as follows.

(1) After merger, the GW spectrum gradually transitions to a phase dominated by fundamental oscillation modes of the remnants (GW ``afterglow''). This is exemplified in our simulations by the initial double-peak structure morphing into a three-peak structure. These modes saturate to a steady level at certain points, putting a check on the initial exponential decay of the GW amplitudes. The result puts forth a first evidence on the emergence of unstable oscillation modes in long-term postmerger GW signals, which may be driven up to potentially detectable levels. It provides validation of BNS mergers as a physical scenario that can excite initially dormant NS unstable oscillation modes. Our results can further motivate future GW observations of high frequencies above 1 kHz.

(2) Template models for long-term postmerger GW signals may require multiple frequency peaks, taking into account double peaks associated with the double-core structure and the new low-frequency peak $f_C$ driven by the CFS instability. It is expected that secondary peaks due to unstable modes may also show up in long-term postmerger evolution with stiff nuclear EOSs and low-mass binaries. Targeted search triggered by inspiral signals can improve the prospect of observing them. A connection with continuous wave search of NS oscillations is also worth considering.

(3) Our simulations suggest a possibility of forming torus-shaped remnants from BNS mergers, which may bear GW signatures different from ellipsoid-shaped remnants. This new qualitative feature adds to the current richness of BNS phenomenology, which will be subject to future observational constraint.

(4) From a dynamical perspective, a toroidal shape substantially delays collapse into BH and prolongs the lifetime of massive metastable remnants. The less compact configuration with relatively low ellipticity, compared with the dumbbell/bar-shaped ones, also gives off persistent GW emission after merger. It sheds light on toroidal models that have been proposed and examined in the context of NS instabilities.

(5) Physical conditions that favor formation of toroidal cores (and double cores as progenitor) include equal-mass binaries, suppressed one-armed instabilities, stiff EOSs, and low masses (equivalently, low compactness). Hence they are less likely to form with softer nuclear EOSs and in higher-mass binaries. Additionally, eccentric binary orbits and NS spins can also contribute to torus formation \cite{Paschalidis15,East16a,East16b}. Various viscous and damping effects may play an important role that deserves further studies (e.g., \cite{Shibata17a,Shibata17b,Radice17}).

(6) Numerical error accumulated over time can readily throw off the phase of waveforms. However, the spectral patterns and properties are less affected. Therefore even simulations with less-than-ideal accuracy can provide reliable trends and take one in the right directions.


\section{Acknowledgement}

We thank He Gao for early discussions that prompted this work, and Mew-Bing Wan for help with the {\tt Einstein Toolkit}. We also thank Yiming Hu, Matthias Hanauske, Enping Zhou, and Tjonnie Li for valuable comments. Our gratitude extends to all the developers of the {\tt Einstein Toolkit} and {\tt LORENE}. The computations were performed on TianHe 2 (NSCC-GZ) at Sun Yat-sen University except for one ($\rmd x=18/64$ in Sec. \ref{sec:resltn}) on LSSC-III (LSEC) at Chinese Academy of Sciences. The work was supported by NSFC grants 91636111, 11690023, 11622546, 11375260, and U1431120. Z. Cao was also supported by ``Fundamental Research Funds for Central Universities.''


\appendix
\section{Hamiltonian Constraint Violation}

\begin{figure}[htbp!]
 \centering
 \includegraphics[width=0.9\textwidth]{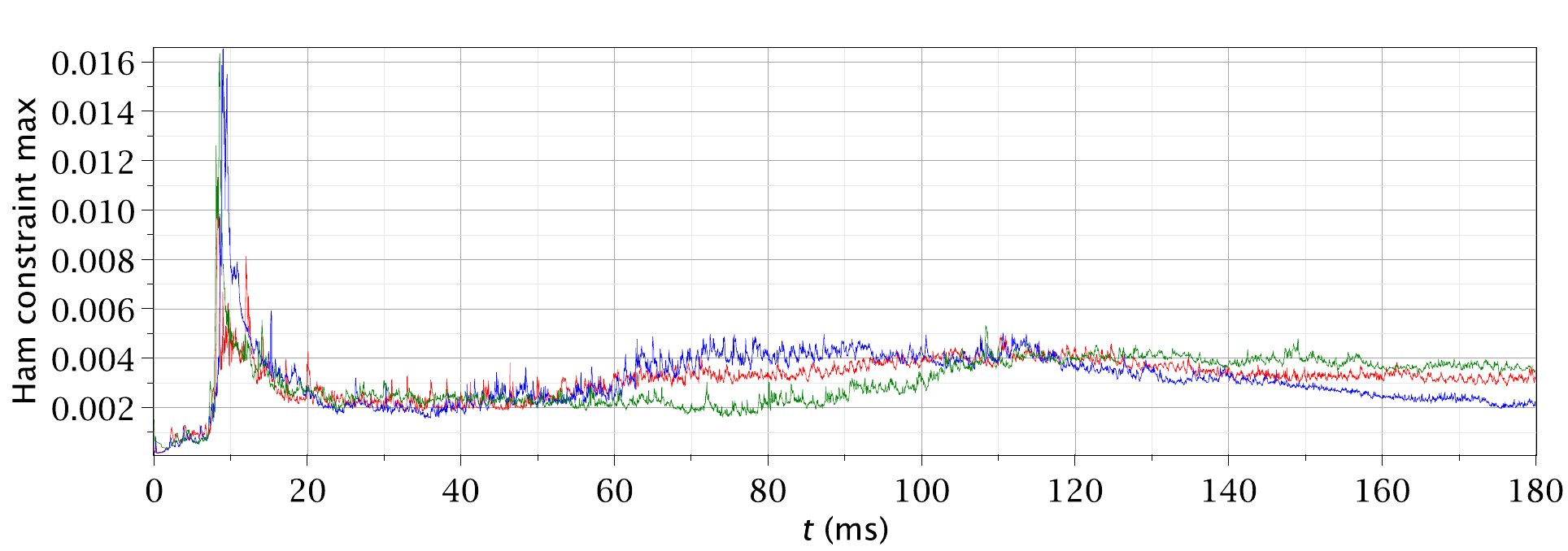}
 \includegraphics[width=0.9\textwidth]{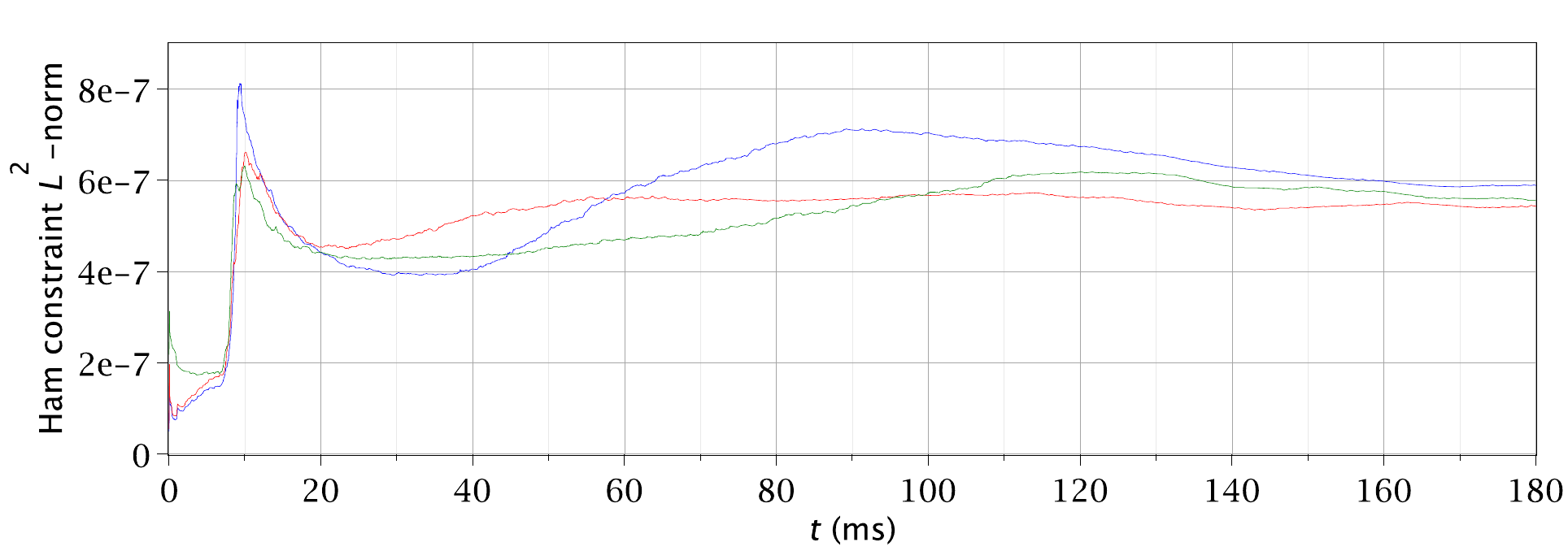}
 \caption{\label{fig_Ham} Time evolution of the maximum and $L^2$-norm of the Hamiltonian constraint violation from the three high-resolution simulations (Sec. \ref{sec:masses}) with NS baryon masses $M_b=1.355$ (green), $1.399$ (blue), $1.445 M_\odot$ (red).}
\end{figure}

In Fig. \ref{fig_Ham}, we monitor violation of the Hamiltonian constraint calculated by the module {\tt ML\_ADMConstraints}. Two plots are shown representing the maximum and the $L^2$-norm. The violation does not increase monotonically, but spikes at the merger ($\sim 10$ ms). Then it drops as the violent dynamics dies down. The $L^2$-norm stays below $10^{-6}$, which is comparable with the accurate evolutions in \cite{Baiotti08,Baiotti09}.

\end{document}